\pdfoutput=1
\documentclass[traditabstract]{aa}
\usepackage{epsfig}
\usepackage{rotating}
\newif\ifAMStwofonts

\def\gtaprx {\lower .1ex\hbox{\rlap{\raise .6ex\hbox{\hskip .3ex
 {\ifmmode{\scriptscriptstyle >}\else {$\scriptscriptstyle >$}\fi}}}
 \kern -.4ex{\ifmmode{\scriptscriptstyle \sim}\else
 {$\scriptscriptstyle\sim$}\fi}}}
\def\ltaprx {\lower .1ex\hbox{\rlap{\raise .6ex\hbox{\hskip .3ex
 {\ifmmode{\scriptscriptstyle <}\else {$\scriptscriptstyle <$}\fi}}}
 \kern -.4ex{\ifmmode{\scriptscriptstyle \sim}\else
 {$\scriptscriptstyle\sim$}\fi}}}
\def\etal {et al. }
\def\littleprime{\ifmmode{\scriptscriptstyle \prime }
 \else{\hbox{$\scriptscriptstyle \prime$ }}\fi}
\def\littless{\ifmmode{\scriptscriptstyle s }
 \else{\hbox{$\scriptscriptstyle s $ }}\fi}
\def\littlemm{\ifmmode{\scriptscriptstyle m }
 \else{\hbox{$\scriptscriptstyle m $ }}\fi}
\def\littlehh{\ifmmode{\scriptscriptstyle h }
 \else{\hbox{$\scriptscriptstyle h $ }}\fi}
\def\littlecirc{\ifmmode{\scriptscriptstyle \circ }
    \else{\hbox{$\scriptscriptstyle \circ $ }}\fi}
\def\rasec{\raise .9ex \hbox{\littless}}
\def\arcsec{\raise .9ex \hbox{\littleprime\hskip-3pt\littleprime\hskip-3pt}}
\def\ramin{\raise .9ex \hbox{\littlemm}}
\def\arcmin{\raise .9ex \hbox{\littleprime}}
\def\hrs{\raise .9ex \hbox{\littlehh}}

\def\degree{\raise .9ex \hbox{\littlecirc}}
\def\magpoint{\hbox to 2pt{}\rlap{\hskip -.5ex \arcmm}.\hbox to 2pt{}}
\def\arcsspoint{\hbox to 1pt{}\rlap{\arcss}.\hbox to 2pt{}}
\def\arcsecpoint{\hbox to 1pt{}\rlap{\arcsec}.\hbox to 2pt{}}
\def\arcminpoint{\hbox to 1pt{}\rlap{\arcmin}.\hbox to 2pt{}}
\def\degreepoint{\hbox to 1pt{}\rlap{\degree}.\hbox to 2pt{}}
\def\lax{{$\mathrel{\hbox{\rlap{\hbox{\lower4pt\hbox{$\sim$}}}\hbox{$<$}}}$}}
\def\gax{{$\mathrel{\hbox{\rlap{\hbox{\lower4pt\hbox{$\sim$}}}\hbox{$>$}}}$}}  
\begin{document}

\title{Optically Faint Radio Sources: Reborn AGN?}

\author{Mercedes E. Filho$^{1}$ \and Jarle Brinchmann$^{1,2}$ \and Catarina Lobo$^{1,3}$ \and Sonia Ant\'on$^{4,5}$}
\institute{Centro de Astrof\'\i sica da Universidade do Porto, Rua das Estrelas, 4150--762 Porto, Portugal 
\and Leiden Observatory, University of Leiden, PO Box 9513, NL--2300 RA Leiden, The Netherlands
\and Departamento de F\'\i sica e Astronomia, Faculdade de Ci\^encias da Universidade do Porto, Rua do Campo Alegre, 
687, 4169--007, Porto, Portugal \and 
Centro de Investiga\c c\~ao em Ci\^encias Geo-Espaciais, Faculdade de Ci\^encias da Universidade do 
Porto, Porto, Portugal \and 
SIM, Faculdade de Ci\^encias da Universidade de Lisboa, Lisboa, Portugal 
}

\date{Accepted 2011.
      Received 2011;
      in original form 2011}


    \abstract {We have discovered a number of relatively strong radio
      sources in the field-of-view of SDSS galaxy clusters which
      present no optical counterparts down to the magnitude limits of
      the SDSS. The optically faint radio sources appear as double-lobed or core-jet
      objects on the FIRST radio images and have projected angular
      sizes ranging from 0.5 to 1.0 arcmin.  We have followed-up these
      sources with near-infrared imaging using the wide-field imager
      HAWK-I on the VLT. $K_s$-band emitting regions, about 1.5 arcsec
      in size and coincident with the centers of the radio structures,
      were detected in all the sources, with magnitudes in the range
      17--20~mag.  We have used spectral modelling to characterize the
      sample sources.  In general, the radio
      properties are similar to those observed in 3CRR sources but the
      optical-radio slopes are consistent with moderate to high
      redshift ($z<$4) gigahertz-peaked spectrum sources. Our results
      suggest that these unusual objects are galaxies whose black hole
      has been recently re-ignited but retain large-scale radio
      structures, signatures of previous AGN activity.} 


\authorrunning{Filho, Brinchmann, Lobo \& Anton}
\titlerunning{Optically Faint Radio Sources: Reborn AGN?}

\maketitle


\section{Introduction}

In the past fifteen years, deep and extensive radio observations of
the Hubble Deep Field (HDF; Richards \etal 1999), and surveys
like the Very Large Array (VLA) 8.4 GHz survey (Fomalont \etal 2002),
{\sc Phoenix} (Hopkins \etal 2003) and {\sc Atlas} (Norris \etal 2006)
have uncovered a number of previously uncatalogued radio sources.
These are characterized by flux densities that range from several
microjansky to hundreds of millijansky and by projected angular sizes
that can be as large as several megaparsec. Cross-correlation studies have
shown that as many as 10-15\% of the compact radio sources have faint
or no optical or infrared counterparts (Hopkins \etal 2003; Sullivan
\etal 2004; Higdon \etal 2005, 2008; Middelberg \etal 2008a; Garn \&
Alexander 2008; Huynh \etal 2010; Norris \etal 2011; Banfield \etal
2011; Zinn \etal 2011; see also Machalski \etal 2001; Rigby \etal
2007). While a significant fraction of the sub-millijansky radio
population appears to be faint star-forming galaxies (e. g. Haarsma
\etal 2000), radio sources that are faint in the optical
or infrared are generally consistent with high redshift ($z>$1),
radio-loud sources or quasars (Huynh \etal 2010; Higdon \etal 2005,
2008; Garn \& Alexander 2008; Jarvis et al. 2009; Norris et al. 2011;
Banfield et al. 2011)

With the aim of studying and establishing the link between active
galactic nuclei (AGN) and cluster properties, we have compiled a
sample of Sloan Digital Sky Survey (SDSS) maxBCG clusters, at redshifts between 0.17 and 0.28,
with radio sources located in projection in their
cores. Surprisingly, we have discovered a number of relatively strong
radio sources in some of the cluster fields that have no optical
counterparts. There are eight such sources, seven of which have
FR II-type (Fanaroff \& Riley 1974) and one of which has a
core-jet-type radio morphology in the Faint Images of the Radio 
Sky at Twenty Centimeter (FIRST; White \etal 1997) maps. These radio sources
possess angular sizes between 0.5--1 arcmin and flux densities between
1--80 mJy. They have not been identified at any other wavelengths
according to our data mining in the NASA IPAC Extragalactic
  Database (NED) and all databases available through the Virtual
  Observatory (VO).

Follow-up deep, near-infrared (NIR) observations were obtained with the wide-field 
imager HAWK-I at the European Southern Observatory (ESO) Very Large Telescope (VLT) in order to try to
identify the radio source host galaxy. In this paper we present the data
and discuss the nature of these objects, mainly based on their spectral energy
distribution (SED) and spectral modelling. 

The paper is organised as follows: Section~2 describes the sample selection, 
Section~3 presents the new observations and data reduction and Section~4 describes the SDSS and Stripe 82 component
identifications. In Section~5 we discuss the spectral energy distribution of the objects and their nature. 
We summarise our conclusions in Section~6. 

Throughout the paper we have assumed the Seven Year Wilkinson Anisotropy Probe (WMAP7) cosmological parameter set --
flat Universe, H$_{\rm 0}$=70 kms$^{-1}$ Mpc$^{-1}$ and $\Omega_m$=0.27 (Larson \etal 2011).

\section{Sample Selection}

The SDSS maxBCG cluster catalog (13~823 clusters; 
Koester \etal 2007) was used as the seed catalogue for our study.
The radio sources were selected in the following manner:

\begin{itemize}
\item We first cross-correlated the cluster sample with the FIRST
 catalogue (White \etal 1997);
\item  We then retained 291 clusters which contained
at least one FIRST object within 1~Mpc in projection from the brightest cluster galaxy (BCG); 
\item Visual inspection of the selected fields provided a subsample of 
radio galaxies with extended double-lobed or core-jet radio morphology; 
\item For these, when a secure SDSS identification was made, the existing 
spectroscopy or multi-band photometry was used to optically characterise the
 source and assess whether the source belongs to the cluster or not.
\end{itemize}

During this process, we have identified eight radio sources with no optical SDSS counterpart, 
which indicates that their $r_{AB}$-band magnitudes must be over 22~mag.
The radio sources are located in the
fields-of-view of clusters with redshifts ranging from 0.17 to 0.28.
The radio sources are further characterized by their radio-loudness 
(as defined by a large radio flux density relative to the optical and NIR; see Section 5.2 and 5.3), 
arcsec-scale FR II-type or core-jet radio morphology and 
relatively strong FIRST flux densities in the range 1 mJy$<$F$_{\rm 1.4\,GHz}<$80 mJy. 

Table~1 contains the radio data for the sample sources. Radio sources are identified by 
the sequential number of the cluster field to which they belong in projection, as it appears in the maxBCG 
cluster catalogue (Koester \etal 2007). Radio source components are identified 
by the position and nature relative to the overall radio source structure. For the only core-jet radio source in our
sample (maxBCG 3131; Fig. 1), we have designated the radio emission regions North Comp and South Comp 1 and 2 (Table~1).
In addition, maxBCG 2596 is the only radio source with a radio core detection (Nucleus; Table~1).
We note that the radio properties of the optically faint radio sources are dominated by the extended 
(lobe) radio emission.

Table~2 contains redshift information on the maxBCG clusters containing our sample sources in their field-of-view. 

\section{Near-Infrared Observations and Data Reduction}

The unidentified radio sources were followed-up with NIR imaging using
the wide-field imager HAWK-I on the ESO UT4 of the VLT. The proposal, with
reference 081.A-0624(A), was awarded a total of 2.7 hours observing time during period 81. 
The total integration time on-source was about 1 000 seconds. The
data were reduced using the HAWK-I data reduction pipeline (version
1.4.2), which includes recipes for darks, flats, zero-point
computation, detector linearity, illumination, additional calibration,
distortion corrections, jittering and image stitching.  The reduced
images were then astrometrically calibrated using the Two Micron
  All Sky Survey (2MASS) catalogue and tools provided in the 
  Graphical Astronomy and Image Analysis Tool (GAIA; version 4.3--0).
The astrometric calibration is good to better than one arcsec,
which is sufficient for our purposes.  Aperture photometry was
performed on the images using the GAIA {\sc se}xtractor program.
$K_s$-band magnitudes (in the Vega magnitude system)
were estimated using zero-points given in the
corresponding quadrant of the standard star images and assuming no
significant color term with respect to 2MASS (ESO HAWK-I Science
Verification, November 7, 2007: Note on
Photometry\footnote{http://www.eso.org/sci/activities/vltsv/hawkisv/}).

Results of the NIR observations are presented in Table~3. NIR sources are identified by 
the sequential number of the cluster field to which they belong in projection, as it appears in the maxBCG 
cluster catalogue (Koester \etal 2007). 
NIR counterparts of the centers of the radio structures are designated as
components "N". Typically the sizes of these central NIR components,
as provided by GAIA {\sc se}xtractor aperture photometry, are
$\sim$1.5 arcsec. For the special case of maxBCG 8495, we have used
GAIA manual aperture photometry to estimate the magnitude in a region
($\sim$2 arcsec) containing the "double nuclear" sources Na and Nb
(component N; Table~3).  NIR sources near or within (in projection)
but not associated with (see Section 5.1) the FIRST radio
sources, are identified 
by the position relative to the overall radio source structure. The quoted NIR magnitudes
are in the Vega magnitude system. 

\section{SDSS and Stripe 82 Data}

The SDSS (York et al. 2000) is a survey covering more than 35\% of the
sky providing deep, photometric observations in 5 bands ($u$, $g$,
$r$, $i$ and $z$; AB magnitude system) of about 500 million objects and spectra for more
than 1 million sources.

We have used the SDSS Data Release 7 (DR7; Abazajian et al. 2009) to
attempt to identify optical counterparts of the radio and NIR
components of our sources. Identification was based on the coincidence
of the positions to less than 1 arcsec, followed by visual inspection
of the images. None of the radio components (Table~1) or NIR emission
regions coincident with the centers of the radio structures
(components N; Table~3) were detected in the SDSS, providing an
$r_{AB}$-band upper limit of $\sim$22~mag.
However, several NIR sources near or within (in projection)
but not associated with (see Section 5.1) the FIRST radio sources, were identified.\\

Stripe 82 is a $\sim$300~deg$^2$ region along the celestial equator,
that has been imaged by a number of different collaborations (with the
same set-up and in the same bands) over one hundred times, providing
coadded optical data two magnitudes deeper than the single-epoch SDSS
data.

Three of our fields are located within the Stripe 82 region, namely
maxBCG 3131, maxBCG 6167 and maxBCG 8495. For these fields, we have
looked for possible optical counterparts to our NIR
components. Identification was based on visual inspection of all
possible counterparts within 5~arcsec of the NIR source. For
each of the three fields, we have detected the faint optical Stripe 82
counterparts of the NIR components coincident with the centers of the
radio structures (components N; Table~3). 
We have also identified the Stripe 82 optical
counterparts of several other NIR sources
near or within (in projection) but not associated with (see Section 5.1) the FIRST
radio sources.

Stripe 82 optical counterparts were also sought directly for the extended radio components. 
In this procedure, Stripe 82 yielded matches within 5~arcsec
of the positions of the radio components in maxBCG 3131 and maxBCG 6167.
For maxBCG 3131, radio components North Comp and South Comp 1 coincide with the position of optical sources; these are rather faint 
($r_{AB}$-band magnitudes of 22.7 and 23.7~mag) extended sources.
As for maxBCG 6167, only its North Lobe radio component flags a cross-match in Stripe 82; this is an extremely faint 
extended object with an ill-determined $r_{AB}$-band magnitude of 24.5$\pm$1.5 mag. Because radio lobes are not 
expected to produce significant optical emission, these cross-matches 
are most likely chance alignments of galaxies unassociated with our radio sources, located in the same line-of-sight of the radio structure.

The faintness of the optical counterparts in the Stripe 82 data prevents us from
deriving structural indicators for the galaxies.

The results of the cross-correlation with the SDSS and Stripe 82 data are included in Table~4.
The optical magnitudes are in the AB magnitude system.

\section{Discussion}

\subsection{Individual Sources}

The FIRST, NIR, SDSS and Stripe 82 (when available) $r_{AB}$-band images of the fields containing
the sample sources are presented in Fig.~1--4. 
HAWK-I observations were successful, as NIR emission, coincident with the centers of the radio structures (components N; Table~3),
was detected in all the targets. 
In the particular case of maxBCG 2596, the central NIR emission source (component N; Table~3) coincides with the radio core 
(Nucleus; Table~1),
while in maxBCG 3131, the central NIR source (component N; Table~3) falls between the radio components North Comp and 
South Comp 1 (Table~1).
We have also investigated all NIR sources near or within (in projection) but not associated with the FIRST radio sources.
In the following, we discuss the HAWK-I sources individually.\\

\noindent {\bf maxBCG 2596} In this object we have identified the NIR counterpart (component N; Table~3) to the radio core (Nucleus; Table~1)
at the center of the radio structure, 
which is the brightest NIR component in our sample, but still falls below the $\sim$22~mag $r_{AB}$-band
limit of the SDSS. We have also detected NIR sources (components W and E; Table~3) coincident with the termination of the radio 
lobes (components East and West Lobe, Table~1), the latter of which display extended NIR morphology. The NIR E and W components are 
detected in the SDSS image, where they are identified with extended, galaxy-type objects -- J010202.12-103258.3 
and J010159.33-103311.2 (Table~4) -- with somewhat similar $r_{AB}$-band magnitudes of $\sim$19.4 and 
$\sim$20.9 (Table~4), but distinct photometric redshifts ($z$=0.09 and 0.19; Table~4). Because radio
lobes do not produce significant amounts of optical emission and due to the redshift discrepancy between the radio source
($z$=0.56; Table~7; see also Section 5.3) and the corresponding optical/NIR components ($z$=0.09 and 0.19; Table~4), we conclude that this
is a chance alignment.
However, according to their redshift, the NIR E component lies in the cluster foreground, whereas the SDSS counterpart 
identified with the NIR W component has a photometric redshift ($z$=0.19; Table~4) consistent with both the photometric redshift 
of the cluster field determined by Koester et al. (2007) and the spectroscopic redshift of the BCG associated 
with the same cluster ($z\sim$0.18; Table~2). One can conclude that, if this cross-identification is correct, then the NIR W component is 
embedded in the cluster and, consistently, has a $g-r$ colour index typical of an early-type galaxy (Fukugita \etal 1995) at the 
cluster distance ($z\sim$0.18; Table~2).\\

\noindent {\bf maxBCG 3131} The HAWKI-I image shows a complex NIR morphology. All of the NIR components 
can be identified on the Stripe 82 images with $r_{AB}$-band magnitudes ranging from 19 to 24~mag (Table~4). The central
NIR source (component N; Table~3), coincident with the center of the radio structure (between North Comp and South Comp 1 components; Table~1)
is the faintest, with a very blue $g-r$=0.35 colour index, consistent with the optical colour of a quasar (Richards \etal 2001) 
at the estimated redshift of our radio source ($z$=1.03; Table~7; see also Section 5.3). 
This result is confirmed by the photometric redshift estimate for the 
Stripe 82 detection ($z$=1.57; Table~7; see also Section 5.3). There is diffuse NIR 
emission to the East (components E1, E2 and E3; Table~3), coincident with the outline of the Northern radio component (component North Comp; Table~1). 
From careful inspection of the Stripe 82 images, it can be seen that components E1 and E2 are
two resolved components of the same edge-on foreground galaxy (relative to the cluster in this field), 
located at a redshift of $z\sim$0.08 (Table~4; as indicated by the 
SDSS photometric redshift of component E2). Component E3 (Table~3) is identified with the $r_{AB}$-band 21.8 mag SDSS 
extended galaxy-type source J003448.99-002130.9 (Table~4), located at a photometric redshift of $z$=0.58 (Table~4), 
higher than the one attributed to the cluster dominating the field-of-view
($z$=0.27; Table~2), but closer than our radio source estimation ($z$=1.03; Table~7; see also Section 5.3). 
Extended NIR galaxy-type emission is also present due South (components S1 and S2; Table~3) of the Southern radio components 
(components South Comp 1 and 2; Table~1) 
and identified with $r_{AB}$-band
$\sim$21-23 mag Stripe 82 sources -- J003447.51-002145.2 and J003447.21-002145.5 (Table~4) -- the former being located at a higher redshift 
($z$=0.46; Table~4) than the one attributed to the cluster ($z$=0.27; Table~2),
but still not coinciding, by far, with the estimated redshift of our radio source ($z$=1.03; Table~7; see also Section 5.3). 
Furthermore, there are points of aligned Southeast-Northwest NIR emission perpendicular to the 
orientation of the radio structure -- components D1 and D2 (Table~3).
There is a $r_{AB}$-band $\sim$23~mag Stripe 82 stellar-like identification of component D1 -- J003448.41-002136.0 (Table~4) -- and a $\sim$21 mag
SDSS identification 
of component D2 -- J003447.71-002126.9 (Table~4) -- an extended galaxy-type object with 
a photometric redshift of $z\sim$0.60 (Table~4), placing it in the background of the cluster ($z$=0.27; Table~2), but at a similar redshift to
component E3 ($z$=0.58; Table~4). 
There is also evidence for faint NIR emission (component L; Table~3) within (in projection) the Northern radio component (component North Comp; Table~1), 
which has been identified with a $r_{AB}$-band $\sim$23 mag stellar-like object in Stripe 82 (Table~4).\\

\noindent {\bf maxBCG 6167} Like its radio emission, the NIR emission associated with this source is faint. 
We detect central NIR emission (component N; Table~3) coincident with the center of the radio structure. An optical counterpart 
has been detected in Stripe 82 -- J002452.65-005201.7 (Table~4) -- as a galaxy-type source with an $r_{AB}$-band magnitude of about 25~mag (Table~4). 
This is the faintest radio, NIR and optical (Stripe 82) source in our sample. Its extreme $g-r$ blue colour of $-$0.44 is affected by large 
magnitude errors 
($\sim$0.4~mag in each of the 2 bands), rendering it too blue, even for the expected values of a quasar (Richards \etal 2001) at the estimated
redshift of our radio source ($z$=1.13; Table~7; see also Section 5.3). This result is confirmed by the photometric redshift estimate for the 
Stripe 82 detection ($z$=1.00; Table~7; see also Section 5.3).\\

\noindent {\bf maxBCG 8495} The NIR image shows a "double nuclear" emission region (components Na and Nb; Table~3) at the center of the radio structure. 
A similarly extended galaxy-type morphology is seen in the 
Stripe 82 image, although Stripe 82 identifies a sole, quite faint, component -- J004359.47+001230.4 (Table~4) -- with an $r_{AB}$-band magnitude of 
about 24~mag and $g-r$=1.31 (Table~4). Given its red colour and the estimated redshift of the radio source ($z$=0.96; Table~7; see also Section 5.3),
we can conclude that the Stripe 82 optical counterpart has a $g-r$ colour index suggestive of a late-type spiral galaxy (Fukugita \etal 1995), 
which is coherent with the disky appearance in the NIR. The red colour generally excludes a quasar host (Richards \etal 2001) at the estimated 
redshift of our radio source ($z$=0.96; Table~7; see also Section 5.3). These results are confirmed by the photometric redshift estimate for the 
Stripe 82 detection ($z$=0.88; Table~7; see also Section 5.3). There is also a small patch of unidentified NIR emission (component D; Table~3) 
near (in projection) the center of the radio source.\\

\noindent {\bf maxBCG 10942} There is a central NIR emission region (component N; Table~3) which is coincident with the center of the radio
structure and some unidentified NIR emission to the West (component W; Table~3) of the Northern radio lobe (component North Lobe, Table~1), in projection.\\

\noindent {\bf maxBCG 11079} This source shows a complex, faint NIR morphology. There is evidence
for central NIR emission (component N; Table~3) coincident with the center of the radio structure, 
which has a disk-like appearance. There are also peaks of unidentified NIR emission 
within (in projection) and/or near the edge of the
Southern (components S1, S2 and S3; Table~3) and Northern radio lobes (component L; Table~3).\\

\noindent {\bf maxBCG 11390} There is a clear NIR detection (component N; Table~3) at the position of the center of the radio structure 
and an unidentified NIR Eastern (component E; Table~3) source.\\

\noindent {\bf maxBCG 11780} There is an unidentified region of NIR emission within (in projection) the Northern radio lobe (component L; Table~3) 
and at the position of the center of the radio structure (component N; Table~3).\\

\noindent All the NIR images reveal central emission, relative to the radio structure (components N; Table~3), that we 
interpret as thermal host-related emission (see Section~5.2). The NIR magnitudes of
these sources range from 17--20 mag (Table~3) and NIR sizes, provided by GAIA {\sc se}xtractor aperture photometry, are typically $\sim$1.5 arcsec.
We further find unassociated (relative to the radio source) NIR emission in seven out of the eight sources. Some have been
identified with stars, some with background or cluster galaxies and still others remain to
be identified (Table~4). With the possible exception of maxBCG 2596 NIR component W (Table~4), none of the optically identified NIR
components are cluster members.

\subsection{Near-Infrared-Radio Correlation}

We present, in Fig.~5 (see also Table~6), a plot of the NIR emission versus the
FIRST flux density. With the exception of maxBCG 8495 component N (Table~3; see Section 3), the plotted NIR flux (in the Vega magnitude system) 
corresponds to GAIA {\sc se}xtractor aperture photometry of the 
NIR components coincident with the
radio structure centers (components N; Table~3). Total radio flux densities were calculated as the sum of the flux densities of all the FIRST radio components (Table~1). We note in passing 
that the extended radio
emission dominates the total radio flux density in all of our sources even when we detect a radio core
(maxBCG 2596) or core-jet (maxBCG 3131) source.
We have estimated a radio "core" flux density using {\sc sao}image ds9 (version 5.1) 
{\sc funtools}\footnote{http://hea-www.harvard.edu/saord/funtools/programs.html} on circular regions of approximately the same size ($\sim$2~arcsec) and centered
on the NIR emission regions N (Table~3). The radio "core" flux density for sources maxBCG 6167, 8495 and 11390 falls below the 3$\times rms$ (root mean
square, as a measure of the noise level) given in the corresponding FIRST images. 

We have run two correlation tests on the data -- the generalized Kendall's $\tau$ 
correlation coefficient estimation (BHK) and 
the correlation probability by Cox's proportional hazard model (Cox-Hazard), as implemented in IRAF (version
2.14.1; Table~5).
Both the BHK and Cox-Hazard tests show a $>$50\% (radio core flux density) and $>$80\% (total radio flux density) probability
that there is no correlation between the FIRST and NIR emission.
At such scales and given the radio morphology, the FIRST radio emission is sampling synchrotron emission 
associated with the AGN. The lack of correlation between the radio and the NIR emission therefore reflects the thermal nature
of the NIR source, likely arising from the host galaxy. This is in contrast to 3CR galaxies (Spinrad 1985), which commonly show non-thermal NIR central regions associated with the AGN 
(Baldi \etal 2010).
 
The infrared-radio correlation, represented by the parameter $q$, is an indicator of the 
dominant emission mechanism in a 
galaxy (Yun \etal 2001); an AGN contribution serves, in general, to lower the value of $q$. 
Empirically it has been shown that starburst-dominated galaxies at redshifts
below 2 typically show a ratio of about $q\sim$1 (Appleton \etal 2004) and $q\sim$2.4 (Yun \etal 2001),
where $q=\log \frac{\rm F_{\rm IR}}{\rm F_{\rm 1.4\,GHz}}$ and IR (infrared) is measured at 24~$\mu$m and a combination of 60 and 100~$\mu$m, 
respectively.
A clear enhancement of the radio emission relative to a pure starburst galaxy
is generally attributed  to the presence of an AGN. We note that 
for typical galaxy starbursts, AGN and composite SEDs (e.g. Huynh \etal 2010), the difference between the 24~$\mu$m 
and 2.2~$\mu$m
flux may be as high as two orders of 
magnitude. Therefore, in the very extreme case, for a starburst-driven galaxy $q\sim-$1.0 in $K_s$-band.

We have proceeded to estimate the infrared-radio correlation 
using the $K_s$-band magnitude of the NIR N components (Table~3) and the radio "core" and total radio flux densities, as defined above. The results
are contained in Table~6. Our sources
show values that are generally consistent with the presence of an AGN ($q<$0),  
with maxBCG 11079 being the most "AGN-dominated" source in the terms of the infrared-radio ratio.

\subsection{The Spectral Energy Distribution}

At first glance, the radio morphologies (Fig.~1--4) 
suggest that our optically faint radio sources may be
similar to the radio sources found in the Third Cambridge Revised
  Catalogue of Radio Sources (3CRR)
catalog\footnote{http://3crr.extragalactic.info/cgi/database}. We
present in Fig.~6--8 plots of the redshift, LLS, radio power and $K$-
(gigahertz-peaked spectrum sources and 3CRR) or $K_s$-band (our
sources) emission. For reasons to be
illustrated below, we also include the gigahertz-peaked spectrum
source (GPS) sample of Labiano \etal (2007). The LLS for the 3CRR
sources have been obtained from the largest angular size (LAS)
measurements and
for the GPS sources the LLS has been obtained from O'Dea \& Baum
(1997), when a measurement is available. The quoted radio powers are not
$k$-corrected (Table~7). Plotted radio powers for the 3CRR are for
the radio cores, while for the
GPS sources these are the total radio powers (Labiano \etal 2007). For
our optically faint radio sources, we use the radio "core" powers
(Table~7) and the total (FIRST) radio powers (Table~7) as
defined in Section 5.2.  For simplicity we have assumed a flat radio
spectrum to normalize the radio powers to 5 GHz. 
$K$-band magnitudes
for the GPS sources were obtained by interpolation for a subset of
these sources with NED NIR photometry. $K$-band magnitudes for the
3CRR sample are emission-line- and aperture-corrected magnitudes for a
subset of 3CRR sources with $K$-band
observations\footnote{https://www.astrosci.ca/users/willottc/kz/3ckz.txt}. $K_s$-band
magnitudes for our sample sources were obtained for
the central NIR components (components N; Table~3) coincident
with the centers of the radio structures.  All $K$- and $K_s$-band
magnitudes are in the Vega system. We expect a small colour term
between $K_s$- and $K$-band: $|K$-$K_s|<$0.1. The redshifts of our
sample sources were obtained as described below.

The GPS source bimodality observed in Fig.~8 (lower right) occurs due to a change
in population; for $z>$1 the only GPS sources with NED NIR photometry
are QSOs, denoting that the GPS population is quite heterogeneous.

The radio source outlier in Fig.~7--8 is maxBCG 11780, the lowest
redshift source in our sample (Table~7). Considering that the FR I/FR II break radio power
increases with the optical luminosity of the host galaxy (Ledlow \& Owen 1996),
then the radio power of maxBXG 11780 puts it near the FR I/FR II threshold.

Overall, if we focus on radio properties alone, we can conclude that our optically faint
radio sources appear similar to the 3CRR sources, grazing the
upper envelope of the 3CRR distribution in terms of radio morphology (Fig.~1--4),
LLS and (total) radio power (Fig.~6--8).

To make further progress, we need to explore the
  radio-to-optical SED. As we have relatively limited multi-wavelength
  information and because the source redshifts are unknown, we adopt a
  similar approach to Huynh \etal (2010): we compare our sample data
  with a sparse set of SED templates with NIR and radio data available
  in NED or in the literature. We have found that our data required a
  considerably larger set of SED templates than that used by Huynh \etal (2010).
We have compared our source SEDs with SED templates from the Third Cambridge Catalogue of Radio Sources (3CR; 169 sources; Spinrad
1985), a GPS sample (32 sources; Labiano \etal 2007) and galaxies
that span a large range of mid-infrared galaxy classifications (Spoon
\etal 2007; their Fig. 1): M82 -- a
star-forming FR I galaxy (class 2C), Arp 220 -- a Seyfert-type 
  Ultraluminous Infrared Galaxy (ULIRG; class 3B), Mrk 231 -- a dusty AGN-dominated (Seyfert 1) ULIRG
(class 1A), Mrk 1501 -- a Seyfert 1.2 flat-spectrum
radio source, 3C 305 -- a Seyfert 2 FR I galaxy, Mrk 668 -- a Seyfert 1.5 GPS source, 3C 273
-- a radio-loud quasar and 3C 295 -- a narrow-line FR II radio galaxy.

Our approach is not to rigorously derive photometric redshift
  estimates, but rather to assess what class of objects our sample
  galaxies are more consistent with and what the implied redshift range
  is. To that end, we commence by comparing the radio-to-$K$ or -$K_s$-band
  luminosity. Fig.~9 shows the results of this exercise. 
The quoted NIR flux  (in the Vega magnitude system) corresponds to that of the central host galaxy 
(components N; Table~3), coincident
with the centers of the radio structures. The radio flux density for our optically faint radio
sources is the total FIRST radio flux density (Table~7) as
defined in Section 5.2, while for the remaining sources, the  
total radio flux density is the 1.4 GHz radio flux density as reported in NED. 
The horizontal dotted lines show the measured flux ratio between the radio
  and NIR for our sample sources. On top of this (solid and dashed lines) we show the
  radio-to-NIR ratio for a range of different sources, with starburst 
  towards the bottom and luminous radio galaxies towards the top. The
  large dots with error bars show the average and 68\% spread around
  the mean for the 3CR (dashed) and GPS (solid) samples.

Fig.~9 clearly shows that the majority of our sample sources possess lower
  radio-to-NIR ratios than the 3CR sample, but much larger than
  starbursts, except perhaps at extremely high redshifts. However, the latter
scenario is ruled out because it would imply unrealistically high
  luminosities in both the radio and optical regime. Viewed as a class,
  our sample sources are more consistent with having characteristics
  similar to that of GPS sources at $z<$4. We stress that the
heterogeniety of the GPS sample does not alter significantly the
results presented in Fig.~9.  Whether we use only the QSOs ($z>$1) or
the entire GPS sample, the result is robust; the median ratio in each
redshift bin of the GPS sample will always fall below the 3CR sample.

For each template SED that the horizontal dotted lines of our sources
  intersect, we can obtain a redshift estimate for our sample sources. 
We do this by first selecting those template SEDs/redshifts
  that reproduce the infrared-radio slope to within 10\% (approximately
  consistent with the error estimates). However, it is obvious from
  Fig.~9 that using the radio-to-NIR ratio alone does not provide a unique
  redshift.  We therefore reject any template SED/redshift solution that would
  predict optical fluxes (SDSS or Stripe 82) brighter than the 1$\sigma$ upper limits. 

The heterogeneity of the SEDs and the limited observational data
  does, however, complicate our attempts to estimate rigorous redshifts
  for our sample sources. We do note that Mrk 668 traces
fairly closely the mean trend of the GPS sources (Fig.~9).
We have therefore adopted this SED template in order to obtain
representative redshifts for our sample sources. This might be misleading for individual targets but it is
  a reasonable approach given that our sample do seem to show typical
  infrared-radio slope values consistent with the mean GPS sample. The
estimated redshifts obtained through this procedure are included in Table~7. We
stress that by using a template GPS source like Mrk 668 provides only
an illustrative estimate of the redshift; there is a degeneracy in the
redshift determination. The most robust aspect is that the very lowest
redshift solutions are excluded when we include the optical upper
limits.

We can obtain an independent check on our redshift estimates
by exploiting the fact that three of the NIR sources (components N corresponding to maxBCG 3131, 6167 and 8485) appear to have 
faint counterparts on the Stripe 82 images. By combining the $u$, $g$, $r$, $i$ and $z$ (AB magnitude system)
photometry from Stripe 82 with our NIR photometry, we have estimated
photometric redshifts using version 2.2 of the Le
Phare\footnote{http://www.cfht.hawaii.edu/$\sim$arnouts/lephare.html} 
photometric redshift code (Arnouts \etal 1999; Ilbert \etal 2006). The
resulting redshift estimates are included in Table~7. As can be
seen, these are in very good agreement with the 
estimates obtained using Mrk 668 as a template.
If we compare these radio source redshift estimates (Table~7) 
with those of the clusters and respective BCGs (Table~2), we can conclude
that, although originally identified in the field-of-view of the clusters, 
the optically faint radio sources in our sample are not members of the SDSS clusters listed in Table~2.

An alternative interpretation for the nature of the optically faint radio sources is that they are predominantly
steep spectrum radio sources, where the steep spectrum 
emission of the lobes dominates the radio flux. This would fit most observational results but the
main obstacle lies in the radio-to-NIR ratio. While there exists a few steep spectrum radio sources 
such as 3C 305 (Fig.~9) that have radio-to-NIR ratios lower or comparable to our sources, 
most have considerably higher ratios. More generally, double-lobed radio quasars 
from the Hough \& Readhead (1989) sample lie, on average, well above our sample in the 
radio-NIR diagram. Even if we use 3C 305 as a template, this results in a large discrepancy between 
the redshift inferred from the radio-to-NIR ratio ($z>$3) and that obtained from the 
Le Phare photometric redshift estimate (Table~7).

The result is that we have large-scale, relatively powerful radio
sources similar in radio properties to 3CRR sources, but whose
optical-radio slopes are consistent with those found in GPS sources.
GPS sources are compact ($\leq$1 kpc), powerful (log P$_{\rm
  1.4\,GHz}\geq$ 25 W Hz$^{-1}$) radio sources with a well-defined
peak ($\sim$1 GHz) in their radio spectra (O'Dea 1998). The spectra is
generally interpreted as being produced by a contained or young radio
source (O'Dea 1998).  

The seemingly contradicting evidence can be
reconciled by considering the so-called "double-double" radio galaxy
scenario (Schoenmakers et al. 2000a; see also Stanghellini \etal
2005). It is well-known that AGN have duty cycles ($\sim$10$^{9-10}$
years), with periods of activity ($\sim$10$^{7}$ years) and periods of
dormency (Franseschini, Vercellone \& Fabian 1998).  Large-scale radio
lobes are produced when the AGN jets deposit energy into intergalactic
medium (IGM) "cocoons". If the nuclear activity is halted, the
luminosity of the radio lobes drops and the radio spectrum becomes
steeper due to radiation and expansion losses. However, such lobe
structures are long-lived, fading away on timescales of about 10$^7$
years (Komissarov \& Gubanov 1994), comparable to the timescale of the
nuclear activity itself.  The re-ignition of the jet-forming activity
can occur via internal instability of the accretion disc (Natarajan \&
Pringle 1998) or gas accretion through an interaction or merger
(Barnes \& Hernquist 1996), potentially producing a GPS spectrum
typical of a young radio source.  If the AGN restart occurs within a
few 10$^7$ years, the large-scale radio lobes will still be visible.

The above theory predicts a large number of such "double-double" radio
sources -- 3C 219 (Clarke \& Burns 1991; Clarke \etal 1992), 4C 26.35
(Owen \& Ledlow 1997), 3C 445 (Kronberg, Wiebelinski \& Graham 1986;
Leahy \etal 1997), 1245+676 (Marecki \etal 2003), B1834+620
(Schoenmakers \etal 2000a, 2000c), B0925+420, B1240+389 and B1450+333
(Schoenmakers \etal 2000a) are just a few examples.  Some of these
sources also show GPS-type spectra -- e.g. 0108+388 (Baum \etal 1990), 
1245+676 (Marecki \etal
2003; Ant\'on \etal 2004; Bondi \etal 2004), B1834+620
(Schoenmakers \etal 2000a, 2000c).  However, GPS sources with
large-scale radio emission appear to be rare (Stanghellini \etal 1990, 2005).

When compared to these classical "double-double" radio sources
(Schoenmaker et al. 2000a), our optically faint radio sources
have, on average, similar projected FIRST LLS and total FIRST radio
powers (Table~7; Fig.~7--8). The similarities occur also with the
3CRR sample in
terms of radio morphology, radio power and FIRST LLS (Table~7;
Fig.~1--4 and 7--8), although our sources appear, on average,
fainter in the $K$-band (Table~3; Fig.~8).  Compared to
the GPS sample (Labiano \etal 2007; see also O'Dea 1998 and
Stanghellini \etal 2005), our sources have, on average, larger
projected FIRST LLS, lower total (FIRST) radio powers (Table~7;
Fig.~7--8) and fainter $K$-band magnitudes (Table~3;
Fig.~8).  

The way to reconcile the GPS-type optical-radio slope with
the large-scale radio structure in our sample
sources is to assume the total FIRST emission 
as an upper limit to the radio power of the hypothetical
GPS. The fact that we do not detect significant radio core emission
in the FIRST images is not deterministic; the moderate redshift of the
sources combined with the low signal-to-noise ratio of the FIRST
images may mask out the presence of such a radio core. In this scenario, 
our sources may be regarded as re-ignited AGN,
with the large-scale structure a relic of a previous cycle of black
hole activity. 

Thus based on the present evidence we would argue that while 
interpreting the optically faint radio sources as steep spectrum 
radio sources would be possible in some cases, overall our sample 
appear to show systematically better agreement with a "double-double" 
radio source scenario in the radio-to-NIR diagram (Fig.~9). 
To robustly distinguish between these two scenarios we do, however, 
require further observational data for our sources. 
Deep, high-resolution radio observations should
assess the presence of small-scale radio structures, signatures of recently
re-instigated AGN activity.

These results are somewhat distinct from previous studies of radio
sources with compact radio morphology that are faint in
the optical or infrared (Higdon \etal 2005, 2008; Garn \& Alexander
2008; Jarvis \etal 2009; Huynh \etal 2010; Norris \etal 2011; Banfield
\etal 2011).  Both the observed infrared-radio correlation and SED
modelling results suggest that these radio sources are mainly
radio-loud AGN at $z>$1 with QSO- or Seyfert 2-type spectra.

\section{Conclusions}
\label{conclusions}

In this paper we have presented new, deep near-infrared images of a
small sample of optically faint radio sources found, in projection,
in low redshift galaxy cluster fields. These new observations have allowed us to detect
near-infrared emitting host galaxies with $K_s \sim$17--20~mag, $\sim$1.5 arcsec in size, 
coincident with the centers of the radio structures. 

We have constructed and analysed the spectral energy distributions of
the sources by comparing them with templates of a large range of
galaxy types. Considering the overall radio properties in isolation,
our sources are rather similar to 3CRR sources. However, when comparing our
NIR fluxes to the radio flux density, we find that our NIR fluxes are
relatively too bright to be consistent with the properties of
classical 3CR sources; rather they show radio-to-optical ratios similar to
those found in GPS sources. The photometric redshifts estimated from the 
radio and NIR data assuming a GPS-like source (Mrk 668) are in good agreement
with optical-NIR photometric redshift estimates, when available.

To reconcile these two observations, we suggest that our sample of
optically faint radio sources present properties that are consistent
with reborn active galactic nuclei that retain, in their large-scale
radio structure, signatures of past black hole activity. If this can
be confirmed, it offers an interesting approach to study duty-cycles
and re-ignition of radio sources.

This scenario can be tested by obtaining deep, high-resolution radio observations of
our sources. Our proposed scheme would predict the existence of compact central
sources similar to GPS sources. Parallel to this it is necessary to obtain deeper
photometric and spectroscopic data to study the stellar populations of the sources and improve their
redshift estimates.



\begin{table*}
\footnotesize
\begin{center}
\begin{minipage}[c]{165mm}
\begin{small}
\caption{{\bf The Radio Data.}
Col. 1: The radio source component corresponding to the SDSS maxBCG cluster field: identification
number is the sequential number (Koester et al. 2007).
Col. 2: NVSS 1.4 GHz, 45\arcsec~resolution radio position (J2000).
Col. 3: NVSS total flux density.
Col. 4: FIRST 1.4 GHz, 5\arcsec~resolution radio position (J2000).
Col. 5: FIRST radio peak flux density.
Col. 6: FIRST integrated radio flux density.
}
\begin{tabular}{l c c c c c}
\hline
\hline 
maxBCG           & RA \hspace{1cm} Dec      & NVSS   &  RA \hspace{1cm} Dec       &  FIRST Peak  &   FIRST Int \\
                 &     (J2000)              & (mJy)  &  (J2000)                   &  (mJy/beam)  &   (mJy)     \\
(1)              &     (2)                  & (3)    &  (4)                       &  (5)         &   (6)       \\
\hline
2596 Total       &  01:02:00.36 -10:33:03.1 & 68.9   & \ldots                     & \ldots       & \ldots \\
2596 Nucleus     &  \ldots                  & \ldots & 01:02:00.498 -10:33:04.95  & 3.80         & 13.94  \\
2596 East Lobe   &  \ldots                  & \ldots & 01:02:01.726 -10:32:58.02  & 9.95         & 19.60  \\
2596 West Lobe   &  \ldots                  & \ldots & 01:01:59.682 -10:33:08.81  & 20.43        & 32.18  \\
                 &                          &        &                            &              &         \\
3131 Total       &  00:34:48.07 -00:21:31.2 & 36.8   & \ldots                     & \ldots       & \ldots  \\
3131 North Comp  & \ldots                   & \ldots & 00:34:48.268 -00:21:27.67  & 16.42        & 24.91  \\
3131 South Comp 1& \ldots                   & \ldots & 00:34:47.665 -00:21:36.57  & 3.75         & 4.21     \\
3131 South Comp 2& \ldots                   & \ldots & 00:34:47.226 -00:21:38.28  & 3.61         & 4.02      \\

                 &                          &        &                            &              &        \\
6167 Total       &  00:24:52.61 -00:52:05.6 & 7.6    & \ldots                     & \ldots       & \ldots \\
6167 North Lobe  & \ldots                   & \ldots & 00:24:52.733 -00:51:54.90  & 2.05         & 3.84  \\
6167 South Lobe  & \ldots                   & \ldots & 00:24:52.332 -00:52:14.76  & 1.92         & 1.87   \\
                 &                          &        &                            &              &        \\
8495 Total       &  00:43:59.54 +00:12:31.2 & 43.5   & \ldots                      & \ldots       & \ldots \\
8495 North Lobe  & \ldots                   & \ldots & 00:44:00.178  +00:12:43.26  & 10.81        & 17.81  \\
8495 South Lobe  & \ldots                   & \ldots & 00:43:58.808  +00:12:16.19  & 9.94         & 14.40 \\
                 &                          &        &                             &              &        \\
10942 Total      &  02:17:57.39 -09:18:22.2 & 75.2   & \ldots                      & \ldots       & \ldots \\
10942 North Lobe & \ldots                   & \ldots & 02:17:57.475  -09:17:54.85  & 7.52         & 15.06  \\
10942 South Lobe & \ldots                   & \ldots & 02:17:57.362  -09:18:28.86  & 26.94        & 50.49  \\
                 &                          &        &                             &              &        \\
11079 Total      &  02:01:28.37 -08:19:55.8 & 173.2  & \ldots                      & \ldots       & \ldots \\
11079 North Lobe & \ldots                   & \ldots & 02:01:28.629 -08:19:47.17   & 72.38        & 86.00  \\
11079 South Lobe & \ldots                   & \ldots & 02:01:28.173 -08:20:05.50   & 61.92        & 77.63 \\
                 &                          &        &                             &              &        \\
11390 Total      &  00:03:28.78 -11:12:55.1 & 10.8   & \ldots                      & \ldots       & \ldots \\
11390 North Lobe & \ldots                   & \ldots & 00:03:28.709  -11:12:49.20  & 4.68         & 5.54   \\
11390 South Lobe & \ldots                   & \ldots & 00:03:28.860  -11:13:02.36  & 3.90         & 4.34   \\
                 &                          &        &                             &              &        \\
11780 Total      &  00:05:57.07 -09:09:01.1 & 21.7   & \ldots                      & \ldots       & \ldots \\
11780 North Lobe & \ldots                   & \ldots & 00:05:57.469  -09:08:48.91  & 3.69         & 9.81   \\
11780 South Lobe & \ldots                   & \ldots & 00:05:56.725  -09:09:08.36  & 8.25         & 12.19  \\

\hline
\end{tabular}
\end{small}
\end{minipage}
\end{center}
\end{table*}

\normalsize



\begin{table*}
\footnotesize
\begin{center}
\begin{minipage}[c]{58mm}
\caption{{\bf SDSS maxBCG Cluster Redshifts.}
Col. 1: SDSS maxBCG cluster field: identification number is the
sequential number (Koester et al. 2007).
Col. 2: The photometric redshift of the SDSS cluster, as determined
for a sample of red sequence member
galaxies (Koester et al. 2007; their Table 1).
Col. 3: The spectroscopic redshift of the brightest cluster galaxy when available
(Koester et al. 2007; their Table 1).
}
\begin{tabular}{l c c}
\hline
\hline
maxBCG          & $z_{\rm photo,\hspace{0.1cm}cluster}$  &  $z_{\rm spec,\hspace{0.1cm}BCG}$ \\
(1)             &    (2)                                 & (3)                        \\

\hline

2596            & 0.18095                                & 0.18850 \\

3131            & 0.27005                                & 0.24860 \\

6167            & 0.16745                                & 0.16269 \\

8495            & 0.22955                                & 0.21876 \\

10942           & 0.26195                                & \ldots \\

11079           & 0.28085                                & \ldots \\

11390           & 0.23495                                & \ldots \\

11780           & 0.20795                                & 0.21456 \\

\hline

\end{tabular}
\end{minipage}
\end{center}
\end{table*}

\normalsize



\begin{table*}
\footnotesize
\begin{center}
\begin{minipage}[c]{118mm}
\caption{{\bf The Near-Infrared Data.}
Col. 1: The NIR source component corresponding to the SDSS maxBCG cluster field: identification
number is the sequential number (Koester et al. 2007).
Col. 2: Near-infrared position (J2000) from the GAIA aperture photometry {\sc se}xtractor fit (except for maxBCG 8495 component N).
Col. 3: Ellipticity of the fitted ellipse (except for maxBCG 8495 component N).
Col. 4: GAIA {\sc se}xtractor aperture photometry $K_s$-band magnitude and error (except for maxBCG 8495 component N). Magnitudes are in the Vega system.
Col. 5: Note on the component: N -- component coincident with the center of the radio structure; SDSS -- SDSS detection;
S82 -- Stripe 82 detection.}

\begin{tabular}{l c c c c}

\hline
\hline

maxBCG       & RA \hspace{1cm}Dec                  & Ellipticity & $K_s$               & Note   \\
             &     (J2000)                         &             & (mag)               &         \\
(1)          &        (2)                          &  (3)        & (4)                 & (5)      \\

\hline

2596 N       & 01:02:00.677 -10:33:02.46           & 0.056       & 17.57$\pm$0.034  & N \\
2596 E       & 01:02:02.089 -10:32:57.99           & 0.045       & 17.22$\pm$0.037  & SDSS \\
2596 W       & 01:01:59.311 -10:33:10.95           & 0.381       & 16.91$\pm$0.023  & SDSS \\
             &                                     &             &                     &    \\
3131 N       & 00:34:48.040 -00:21:32.47           & 0.432       & 19.21$\pm$0.117  & N/S82 \\
3131 D1      & 00:34:48.421 -00:21:35.99           & 0.128       & 17.78$\pm$0.040  & S82 \\
3131 D2      & 00:34:47.719 -00:21:27.19           & 0.208       & 17.51$\pm$0.034  & SDSS/S82 \\
3131 L       & 00:34:48.333 -00:21:29.70           & 0.540       & 19.59$\pm$0.143  & S82  \\
3131 E1$^a$  & 00:34:48.774 -00:21:20.20           & 0.843       & 19.62$\pm$0.164  & S82 \\
3131 E2$^a$  & 00:34:48.954 -00:21:25.16           & 0.219       & 18.97$\pm$0.097  & SDSS/S82 \\
3131 E3      & 00:34:48.990 -00:21:31.22           & 0.097       & 17.61$\pm$0.036  & SDSS/S82 \\
3131 S1      & 00:34:47.521 -00:21:45.26           & 0.426       & 17.66$\pm$0.037  & SDSS/S82 \\
3131 S2      & 00:34:47.211 -00:21:45.72           & 0.125       & 18.23$\pm$0.055  & S82  \\
             &                                     &             &                     &       \\
6167 N       & 00:24:52.655 -00:52:01.49           & 0.073       & 19.88$\pm$0.149  & N/S82 \\
             &                                     &             &                     &         \\
8495 Na$^b$  & 00:43:59.447 +00:12:30.01           & 0.237       & 18.20$\pm$0.072  & \ldots \\
8495 Nb$^b$  & 00:43:59.526 +00:12:30.84           & 0.256       & 18.56$\pm$0.096  & \ldots \\
8495 N$^c$   & 00:43:59.488 +00:12:30.60           & 0.880       & 17.47$\pm$0.048  & N/S82 \\      
8495 D       & 00:43:59.498 +00:12:26.22           & 0.423       & 18.84$\pm$0.121  & \ldots \\
             &                                     &             &                     &       \\
10942 N      & 02:17:57.382 -09:18:18.10           & 0.053       & 18.03$\pm$0.048  & N  \\
10942 W      & 02:17:57.166 -09:18:05.27           & 0.147       & 19.05$\pm$0.096  & \ldots \\
             &                                     &             &                     &         \\
11079 N      & 02:01:28.258 -08:19:56.63           & 0.303       & 19.16$\pm$0.112  & N \\
11079 S1     & 02:01:28.433 -08:20:01.79           & 0.398       & 19.10$\pm$0.108  & \ldots \\
11079 S2     & 02:01:28.493 -08:20:11.35           & 0.275       & 19.38$\pm$0.135  & \ldots \\
11079 S3     & 02:01:28.691 -08:20:08.10           & 0.444       & 19.24$\pm$0.120  & \ldots \\
11079 L      & 02:01:28.865 -08:19:46.39           & 0.051       & 18.92$\pm$0.093  & \ldots \\
             &                                     &             &                     &       \\
11390 N      & 00:03:28.911 -11:12:53.63           & 0.101       & 19.24$\pm$0.109  & N \\
11390 E      & 00:03:29.109 -11:12:52.75           & 0.365       & 19.44$\pm$0.128  & \ldots \\
             &                                     &             &                     &        \\
11780 N      & 00:05:57.061 -09:08:58.43           & 0.059       & 18.06$\pm$0.040  & N \\
11780 L      & 00:05:57.324 -09:08:52.20           & 0.051       & 19.29$\pm$0.099  & \ldots \\

\hline
\end{tabular}

$^{a}$ Stripe 82 images show that components E1 and E2 are two resolved components of the same edge-on galaxy.
 
$^{b}$ The maxBCG 8495 Na and Nb components refer to the "double nuclear" emission region detected in the $K_s$-band.

$^{c}$ The maxBCG 8495 N component refers to GAIA manual aperture photometry of a region ($\sim$2 arcsec) containing the "double nuclear" components Na and Nb.

\end{minipage}
\end{center}
\end{table*}

\normalsize


%

\begin{table*}
\scriptsize
\begin{center}
\begin{minipage}[c]{177mm}
\caption{{\bf The Optical Data.} Radio (top) or NIR (bottom) component positions.
Col. 1: SDSS maxBCG cluster field and radio or NIR component: identification number
is the sequential number (Koester et al. 2007).
Col. 2: FIRST or NIR component position (J2000).
Col. 3: SDSS or Stripe 82 (S82) detection.
Col. 4: IAU designation of the SDSS or Stripe 82 counterpart.
Col. 5: Separation between the position of our radio or NIR component and the SDSS or Stripe 82 detection.
Col. 6: SDSS or Stripe 82 $r_{AB}$-band (model) magnitude, corrected for
Galactic extinction, and respective error. Magnitudes are in the AB system.
Col. 7: $g-r$ colour index.
Col. 8: Photometric redshift and error (only available for SDSS detections).
Col. 9: Type of SDSS or Stripe 82 source: G -- galaxy; S -- star.}
\label{sdssdata}

\begin{tabular}{l c c c c c c c c}

\hline
\hline

maxBCG           &     RA \hspace{1cm} Dec    &  Survey         & IAU ID           & $\theta$  & $r_{AB}$              & $g-r$              & $z_{\rm phot}$             & Source \\
                 &     (J2000)                &                 &                  & (")  & (mag)            &(mag)               &                            & type  \\
 
   (1)           &       (2)                  & (3)             & (4)              & (5)              & (6)             & (7)    & (8) & (9)  \\

\hline

2596 Nucleus     & 01:02:00.498 -10:33:04.95  & \ldots          & \ldots         & \ldots           & \ldots           &           \ldots          & \ldots          & \ldots  \\
2596 East Lobe   & 01:02:01.726 -10:32:58.02  & \ldots          & \ldots         & \ldots           & \ldots           &          \ldots          & \ldots          & \ldots \\
2596 West Lobe   & 01:01:59.682 -10:33:08.81  & \ldots          & \ldots         & \ldots           & \ldots           &          \ldots          & \ldots          & \ldots  \\
                 &                            &                 &                &                  &                 &        &        \\
3131 North Comp$^a$  & 00:34:48.268 -00:21:27.67  & S82             & J003448.31-002128.9    & 1.48 & 22.74$\pm$0.35   & 0.89 & \ldots          & G       \\
3131 South Comp 1$^a$& 00:34:47.665 -00:21:36.57  & S82             & J003447.44-002136.2    & 3.28 & 23.67$\pm$0.15   & 1.06 & \ldots          & G       \\
3131 South Comp 2& 00:34:47.226 -00:21:38.28  & \ldots          & \ldots         & \ldots           & \ldots           &       \ldots & \ldots          & \ldots \\
                 &                            &                 &                &                  &                 &        &        \\
6167 North Lobe$^a$  & 00:24:52.733 -00:51:54.90  & S82             & J002452.50-005153.1    & 3.87 & 24.54$\pm$1.46   & 0.87 & \ldots          & G  \\
6167 South Lobe  & 00:24:52.332 -00:52:14.76  & \ldots          & \ldots         & \ldots           & \ldots           & \ldots          & \ldots          & \ldots  \\
                 &                            &                 &                &                  &                 &        &       \\
8495 North Lobe  & 00:44:00.178 +00:12:43.26  & \ldots          & \ldots          & \ldots           & \ldots           & \ldots          & \ldots          & \ldots  \\
8495 South Lobe  & 00:43:58.808 +00:12:16.19  & \ldots          & \ldots          & \ldots           & \ldots           & \ldots          & \ldots          & \ldots \\
                 &                            &                 &                 &                  &                 &        & \\
10942 North Lobe & 02:17:57.475 -09:17:54.85  & \ldots          & \ldots          & \ldots           & \ldots           & \ldots          & \ldots          & \ldots  \\
10942 South Lobe & 02:17:57.362 -09:18:28.86  & \ldots          & \ldots          & \ldots           & \ldots           & \ldots          & \ldots          & \ldots  \\
                 &                            &                 &                 &                  &                 &        &    \\
11079 North Lobe & 02:01:28.629 -08:19:47.17  & \ldots          & \ldots          & \ldots           & \ldots           & \ldots          & \ldots          & \ldots  \\
11079 South Lobe & 02:01:28.173 -08:20:05.50  & \ldots          & \ldots          & \ldots           & \ldots           & \ldots          & \ldots          & \ldots  \\
                 &                            &                 &                 &                  &                 &        & \\
11390 North Lobe & 00:03:28.709 -11:12:49.20  & \ldots          & \ldots          & \ldots           & \ldots           & \ldots          & \ldots          & \ldots  \\
11390 South Lobe & 00:03:28.860 -11:13:02.36  & \ldots          & \ldots          & \ldots           & \ldots           & \ldots          & \ldots          & \ldots  \\
                 &                            &                 &                 &                  &                 &        &  \\
11780 North Lobe & 00:05:57.469 -09:08:48.91  & \ldots          & \ldots          & \ldots           & \ldots           & \ldots          & \ldots          & \ldots  \\
11780 South Lobe & 00:05:56.725 -09:09:08.36  & \ldots          & \ldots          & \ldots           & \ldots           & \ldots          & \ldots          & \ldots  \\

\hline

2596 N       & 01:02:00.677 -10:33:02.46 & \ldots   & \ldots              & \ldots         & \ldots           & \ldots                & \ldots  & \ldots  \\
2596 E       & 01:02:02.089 -10:32:57.99 & SDSS     & J010202.12-103258.3 & 0.61 & 19.41$\pm$0.03 & 0.76 & 0.09$\pm$0.05 & G      \\
2596 W       & 01:01:59.311 -10:33:10.95 & SDSS     & J010159.33-103311.2 & 0.44 & 20.94$\pm$0.11 & 1.10 & 0.19$\pm$0.06 & G      \\
& & & & & & & \\
3131 N       & 00:34:48.040 -00:21:32.47 & S82      & J003448.15-002131.5 & 1.98 & 24.10$\pm$0.23 & 0.35 & \ldots                & G      \\
3131 D1      & 00:34:48.421 -00:21:35.99 & S82      & J003448.41-002136.0 & 0.11 & 23.03$\pm$0.29 & 1.85 & \ldots                & S       \\
3131 D2      & 00:34:47.719 -00:21:27.19 & SDSS     & J003447.71-002126.9 & 0.29 & 21.39$\pm$0.12 & 1.31 & 0.60$\pm$0.04 & G         \\
             & 00:34:47.719 -00:21:27.19 & S82      & J003447.72-002127.2 & 0.05 & 21.41$\pm$0.11 & 1.72 & \ldots                & G      \\
3131 L       & 00:34:48.333 -00:21:29.70 & S82      & J003448.32-002129.6 & 0.14 & 23.24$\pm$0.27 & 0.00 & \ldots                & S       \\
3131 E1$^b$  & 00:34:48.774 -00:21:20.20 & S82      & J003448.84+002120.7 & 1.24 & 23.02$\pm$0.45 & 0.56 & \ldots                & G       \\
3131 E2$^b$  & 00:34:48.954 -00:21:25.16 & SDSS     & J003448.94-002124.6 & 0.56 & 19.25$\pm$0.04 & 0.44 & 0.08$\pm$0.02 & G        \\
             & 00:34:48.954 -00:21:25.16 & S82      & J003448.95-002125.0 & 0.14 & 19.11$\pm$0.04 & 0.33 & \ldots                & G      \\
3131 E3      & 00:34:48.990 -00:21:31.22 & SDSS     & J003448.99-002130.9 & 0.30 & 21.81$\pm$0.11 & 0.93 & 0.58$\pm$0.13 & G      \\
             & 00:34:48.990 -00:21:31.22 & S82      & J003448.99-002131.1 & 0.10 & 21.48$\pm$0.12 & 2.15 & \ldots                & G      \\
3131 S1      & 00:34:47.521 -00:21:45.26 & SDSS     & J003447.50-002145.3 & 0.26 & 20.63$\pm$0.10 & 1.20 & 0.46$\pm$0.13 & G       \\
             & 00:34:47.521 -00:21:45.26 & S82      & J003447.51-002145.2 & 0.06 & 20.81$\pm$0.10 & 0.98 & \ldots                & G       \\
3131 S2      & 00:34:47.211 -00:21:45.72 & S82      & J003447.21-002145.5 & 0.15 & 22.54$\pm$0.21 & 1.71 & \ldots                & G       \\
& & & & & & & \\
6167 N       & 00:24:52.655 -00:52:01.49 & S82      & J002452.65-005201.7 & 0.22 & 24.88$\pm$0.42 & -0.44 & \ldots                & G      \\
& & & & & & & \\
8495 N$^c$   & 00:43:59.447 +00:12:30.01 & S82     & J004359.47+001230.4 & 0.61 & 23.82$\pm$0.19 & 1.31 & \ldots                & G        \\
8495 D        & 00:43:59.498 +00:12:26.22 & \ldots  & \ldots              & \ldots         & \ldots           & \ldots                & \ldots   & \ldots   \\
& & & & & & & \\
10942 N       & 02:17:57.382 -09:18:18.10 & \ldots  & \ldots              & \ldots         & \ldots           & \ldots                & \ldots   & \ldots   \\
10942 W       & 02:17:57.166 -09:18:05.27 & \ldots  & \ldots              & \ldots         & \ldots           & \ldots                & \ldots   & \ldots   \\
& & & & & & & \\
11079 N       & 02:01:28.258 -08:19:56.63 & \ldots  & \ldots              & \ldots         & \ldots           & \ldots                & \ldots   & \ldots   \\
11079 S1      & 02:01:28.433 -08:20:01.79 & \ldots  & \ldots              & \ldots         & \ldots           & \ldots                & \ldots   & \ldots   \\
11079 S2      & 02:01:28.493 -08:20:11.35 & \ldots  & \ldots              & \ldots         & \ldots           & \ldots                & \ldots   & \ldots   \\
11079 S3      & 02:01:28.691 -08:20:08.10 & \ldots  & \ldots              & \ldots         & \ldots           & \ldots                & \ldots   & \ldots   \\
11079 L       & 02:01:28.865 -08:19:46.39 & \ldots  & \ldots              & \ldots         & \ldots           & \ldots                & \ldots   & \ldots   \\
& & & & & & & \\
11390 N       & 00:03:28.911 -11:12:53.63 & \ldots  & \ldots              & \ldots         & \ldots           & \ldots                & \ldots   & \ldots   \\
11390 E       & 00:03:29.109 -11:12:52.75 & \ldots  & \ldots              & \ldots         & \ldots           & \ldots                & \ldots   & \ldots   \\
& & & & & & & \\
11780 N       & 00:05:57.061 -09:08:58.43 & \ldots  & \ldots              & \ldots         & \ldots           & \ldots                & \ldots   & \ldots   \\
11780 L       & 00:05:57.324 -09:08:52.20 & \ldots  & \ldots              & \ldots         & \ldots           & \ldots                & \ldots   & \ldots   \\

\hline

\end{tabular}

$^{a}$ The Stripe 82 identification of an optical source at the position of the radio component is a chance alignment.

$^{b}$ Stripe 82 images show that components E1 and E2 are two resolved components of the same edge-on galaxy.

$^{c}$ The maxBCG 8495 N component refers to GAIA manual aperture photometry of a region ($\sim$2 arcsec) containing the "double nuclear" components Na and Nb.

\end{minipage}

\end{center}

\end{table*}

\normalsize



\begin{table*}

\footnotesize

\begin{center}

\begin{minipage}[c]{94mm}

\caption{{\bf Correlation Statistics.}
Col. 1: Statistical correlation method: BHK -- generalized  Kendall's
$\tau$ correlation coefficient; Cox -- correlation
probability by Cox's proportional hazard model.
Col. 2 and 3: Independent and dependent variable: NIR pertains to the $K_s$-band fluxes of the NIR components (components N) 
coincident with the radio structure centers (except for maxBCG 8495); FIRST Tot pertains to the FIRST total radio as the sum of the flux density of all the FIRST components;
FIRST Core pertains to the radio "core" flux density estimated using {\sc funtools} 
on circular regions of approximately the same size ($\sim$2~arcsec) and centered
on the NIR emission regions N. 
Col. 4: Statistical results. The test assumes the null hypotheses:
"prob" is the probability that there is no
correlation between the variables.
}

\begin{tabular}{l c c c}

\hline
\hline

Method   &       Variable   & Variable      & Result  \\

(1)      &        (2)       &  (3)          &   (4)   \\

\hline

BHK      & FIRST Tot        & NIR           & Kendall's $\tau$ =  0.7143 \\
         &                  &               & Z value =  1.237 \\
         &                  &               & prob = 0.2160 \\
         & FIRST Core       & NIR           & Kendall's $\tau$ =  0.4000 \\
         &                  &               & Z value =  0.490 \\
         &                  &               & prob =  0.6242 \\
         & NIR              & FIRST Tot     & Kendall's $\tau$ =  0.7143 \\
         &                  &               & Z value =  1.237 \\
         &                  &               & prob =  0.2160 \\
         & NIR              & FIRST Core    & Kendall's $\tau$ =  0.4000 \\
         &                  &               & Z value =  0.490 \\
         &                  &               & prob =  0.6242 \\

\hline

Cox      & FIRST Tot        & NIR           & $\chi^2$ = 0.036 \\
         &                  &               & prob = 0.8491 \\
         & FIRST Core       & NIR           & $\chi^2$ = 0.319 \\
         &                  &               & prob = 0.5724 \\
Cox      & NIR              & FIRST Tot     & $\chi^2$ =  0.052 \\
         &                  &               & prob =  0.8192 \\
         & NIR              & FIRST Core    & $\chi^2$ =  0.065 \\
         &                  &               & prob =  0.7990 \\

\hline

\end{tabular}

\end{minipage}

\end{center}

\end{table*}

\normalsize



\begin{table*}

\footnotesize

\begin{center}

\begin{minipage}[c]{130mm}

\caption{{\bf The Infrared-Radio Correlation.}
Col. 1: The radio and NIR source (components N) corresponding to the SDSS maxBCG cluster field: identification number is the
sequential number (Koester et al. 2007).
Col. 2: NIR magnitude and error of the central components N, from GAIA {\sc se}xtractor aperture photometry (except for maxBCG 8495). 
Magnitudes are in the Vega system.
Col. 3: NIR flux and error of the central components N, from GAIA {\sc se}xtractor aperture photometry (except for maxBCG 8495).
Col. 4: FIRST radio "core" flux density estimated using {\sc funtools} 
on circular regions of approximately the same size ($\sim$2~arcsec) and centered
on the NIR emission regions N and the root mean square, as a measure of the
noise, in the FIRST image.
Col. 5: FIRST total radio flux density estimated from sum of the integrated flux densities of all the FIRST components.
Col. 6: The logarithm and error of the NIR to radio "core" flux density.
Col. 7: The logarithm and error of the NIR to total radio flux density.}

\begin{tabular}{l c c c c c c}

\hline
\hline

maxBCG  &      NIR                &    NIR          &   FIRST Core & First Tot  &   $q_{\rm core}$ &  $q_{\rm tot}$   \\
        &      (mag)              &   (mJy)         &   (mJy)     & (mJy)       &                 & \\
(1)     &      (2)                &   (3)           &   (4)       &  (5)        & (6)             & (7)  \\

\hline

2596    & 17.57$\pm$0.034      & 0.0617$\pm$0.0019 &  1.25$\pm$0.16 &  65.72$\pm$0.16 & -1.31  & -3.03 \\
3131    & 19.21$\pm$0.117      & 0.0114$\pm$0.0015 &  0.91$\pm$0.16 &  33.14$\pm$0.16 & -1.90  & -3.46 \\
6167    & 19.88$\pm$0.149      & 0.0074$\pm$0.0010 &  \ldots        &   5.71$\pm$0.13 & \ldots  & -2.89 \\
8495$^a$& 17.47$\pm$0.044      & 0.0681$\pm$0.0028 &  \ldots        &  32.21$\pm$0.13 & \ldots & -2.67 \\
10942   & 18.03$\pm$0.072      & 0.0404$\pm$0.0027 &  1.01$\pm$0.15 &  65.55$\pm$0.15 & -1.40  & -3.21 \\
11079   & 19.16$\pm$0.112      & 0.0144$\pm$0.0015 &  1.32$\pm$0.13 & 163.63$\pm$0.13 & -1.96  & -4.10 \\
11390   & 19.25$\pm$0.109      & 0.0133$\pm$0.0013 &  \ldots        &   9.88$\pm$0.13 & \ldots  & -2.87 \\ 
11780   & 18.06$\pm$0.099      & 0.0393$\pm$0.0036 &  0.39$\pm$0.13 &   22.0$\pm$0.13 & -1.00  & -2.75 \\

\hline

\end{tabular}

$^{a}$ The maxBCG 8495 NIR magnitude and flux refer to GAIA manual aperture photometry of a region ($\sim$2 arcsec) containing the "double nuclear" components Na and Nb.

\end{minipage}

\end{center}

\end{table*}



\begin{table*}

\footnotesize

\begin{center}

\begin{minipage}[c]{125mm}

\caption{{\bf Redshift and Redshift-Dependent Parameters.}
Col. 1: The radio source corresponding to the SDSS maxBCG cluster field: identification number is the
sequential number (Koester et al. 2007).
Col. 2: Redshift estimate assuming Mrk 668 as a template.
Col. 3: Redshift estimate using the Le Phare photometric redshift code and Stripe 82 photometry in conjunction with our NIR data.
Col. 4: The logarithm of the "core" rest-frame monochromatic radio luminosity. The radio "core" power is estimated using {\sc funtools} 
on circular regions of approximately the same size ($\sim$2~arcsec) and centered
on the NIR emission regions N.  We have not included a $k$-correction.
Col. 5: The logarithm of the total FIRST rest-frame monochromatic radio luminosity. The radio power is estimated from
sum of the integrated flux densities of all the FIRST components. We have not included a $k$-correction.
Col. 6: FIRST projected largest angular size.
Col. 7: FIRST projected largest linear size.
 }

\begin{tabular}{l c c c c c c c}

\hline
\hline

maxBCG         &        $z_{\mathrm{Mrk 668}}$ & $z_{\mathrm{Le\,Phare}}$ & log P$_{\rm core,\,1.4\,GHz}$                  & log P$_{\rm tot,\,1.4\,GHz}$     & LAS       & LLS \\
               &                              &                        & (W Hz$^{-1}$)                          	     & (W Hz$^{-1}$)         	 	& (")        & (Mpc) \\
(1)            &       (2)                    & (3)                    & (4)					     & (5)			        & (6)       & (7)  \\

\hline

2596           &    0.56 &  \ldots       	    & 24.21      & 25.93 & 42 & 0.27 \\ 
3131           &    1.03 &  1.57$^{+1.14}_{-0.43}$  & 24.73      & 26.29 & 27 & 0.22 \\ 
6167           &    1.13 &  1.00$^{+3.81}_{-0.40}$  & \ldots     & 26.63 & 36 & 0.30 \\ 
8495           &    0.96 &  0.88$^{+0.09}_{-0.08}$  & \ldots     & 26.20 & 39 & 0.31 \\   
10942          &    1.71 &  \ldots                   &  25.32     & 27.14 & 51 & 0.44 \\ 
11079          &    3.18 &  \ldots                   & 26.10      & 28.19 & 39 & 0.30 \\
11390          &    1.28 &  \ldots                   & \ldots     & 26.00 & 24 & 0.21 \\ 
11780          &    0.05 &  \ldots                   & 21.37      & 23.12 & 36 & 0.04 \\ 

\hline

\end{tabular}
\end{minipage}

\end{center}

\end{table*}


\clearpage


\begin{figure*}
\centering
\includegraphics[width=8cm]{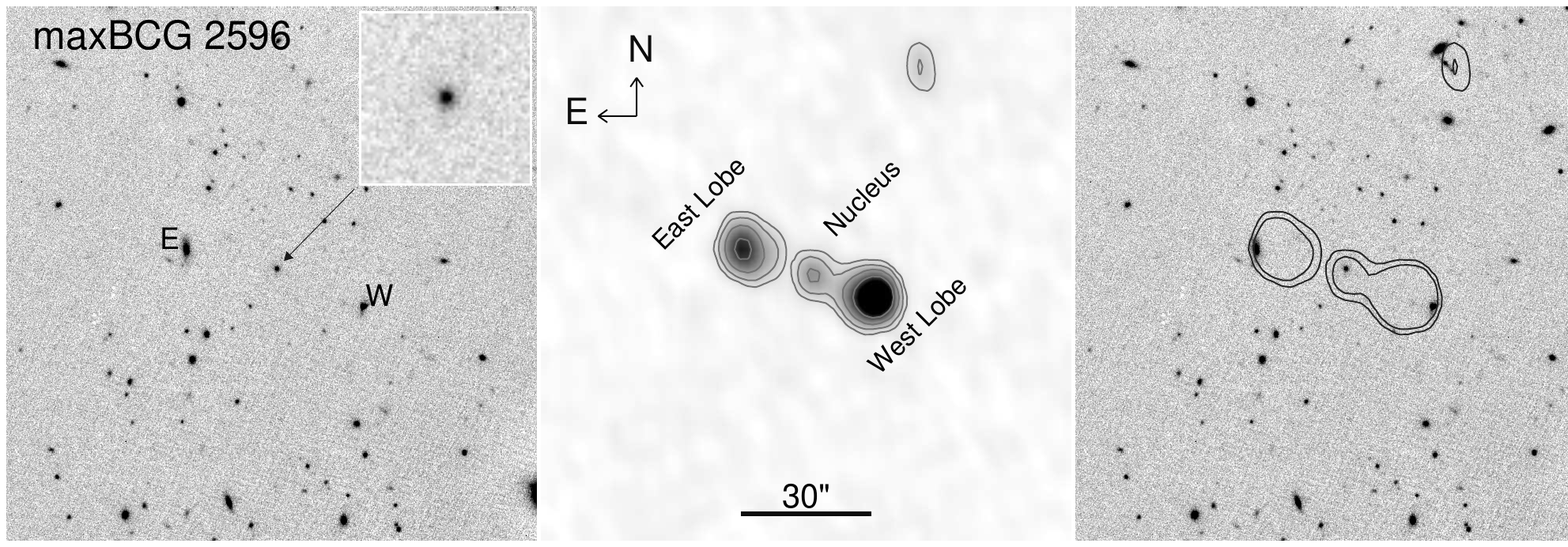}

\includegraphics[width=8cm]{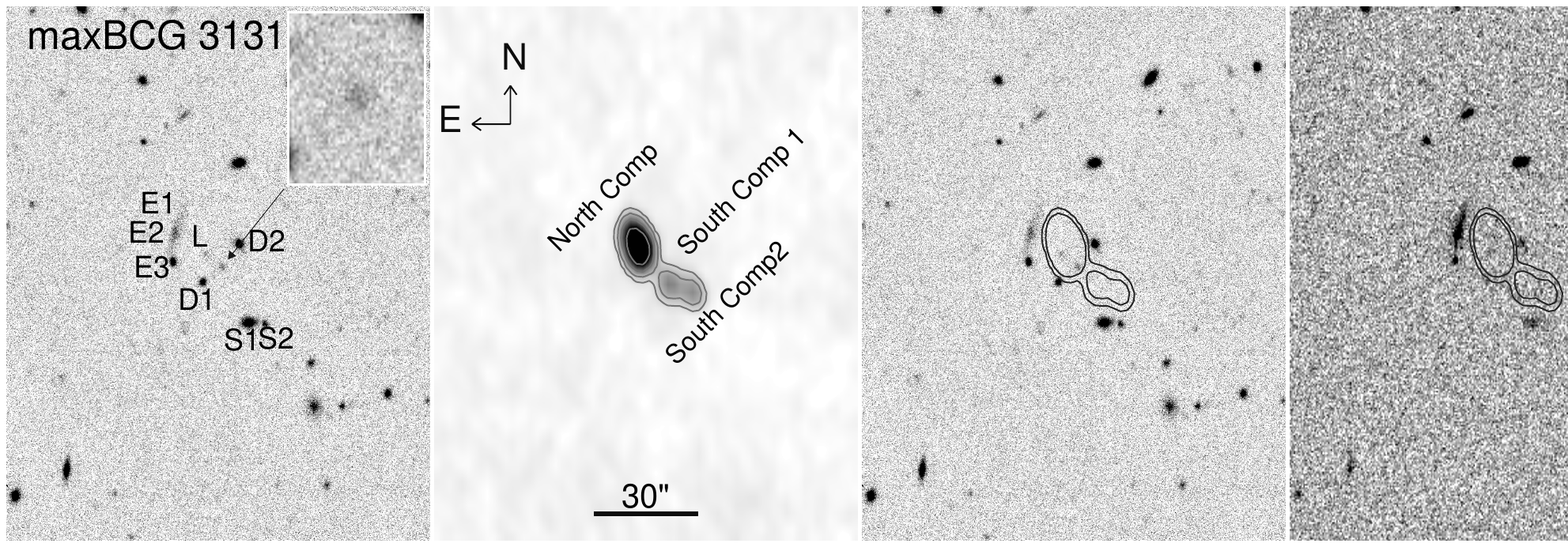}

\caption{From left to right, inverted greyscale image of the NIR
emission, FIRST greyscale emission with superimposed radio contours,
the first two levels of the FIRST contour plots superimposed on the NIR field and
likewise superimposed on the SDSS 
and the Stripe 82 (maxBCG 3131) $r_{AB}$-band field. The field numbers are the sequential numbers of SDSS macxBCG cluster
fields to which the radio sources belong, as they appear in Koester et al. (2007), letters/numbers are
the NIR components and lobe designations refer to the radio components. The inlays contain the NIR N component (Table~3 and 4), coincident
with the center of the radio structure.
Radio contours are: (top) 0.001, 0.002, 0.004, 0.008
mJy; (bottom) 0.001, 0.002, 0.004, 0.008 mJy.}
\label{2596_3131}
\end{figure*}
\normalsize


\clearpage


\begin{figure*}
\centering
\includegraphics[width=8cm]{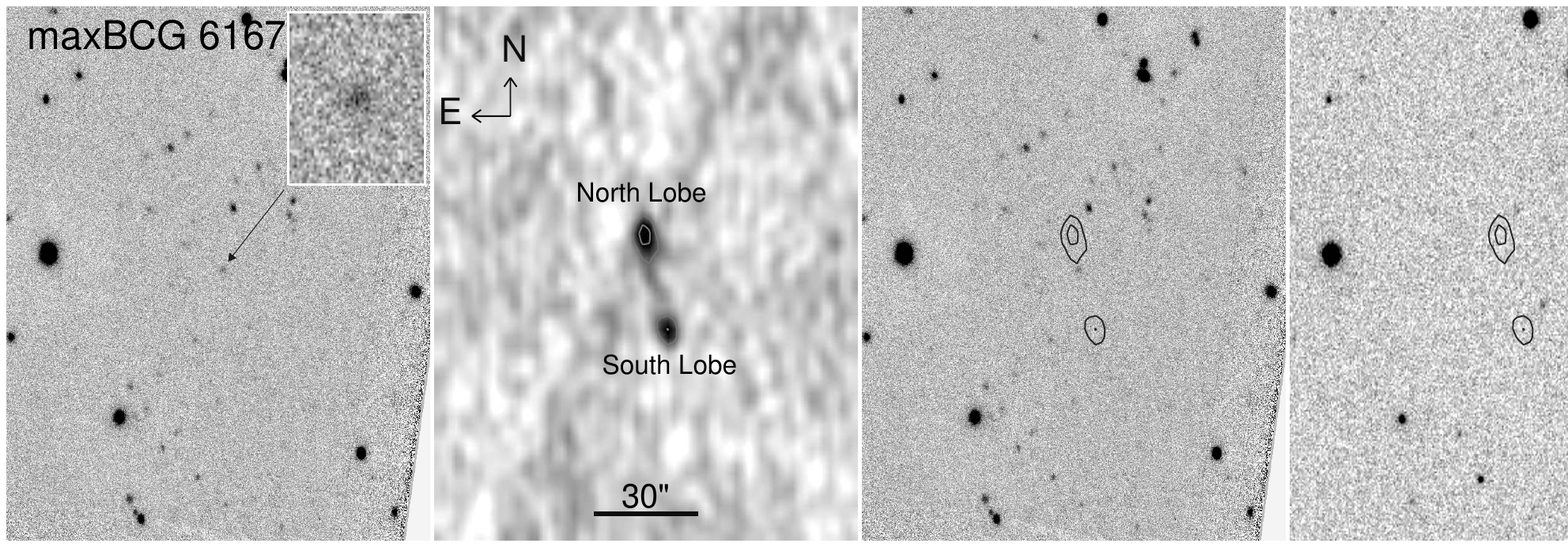}

\includegraphics[width=8cm]{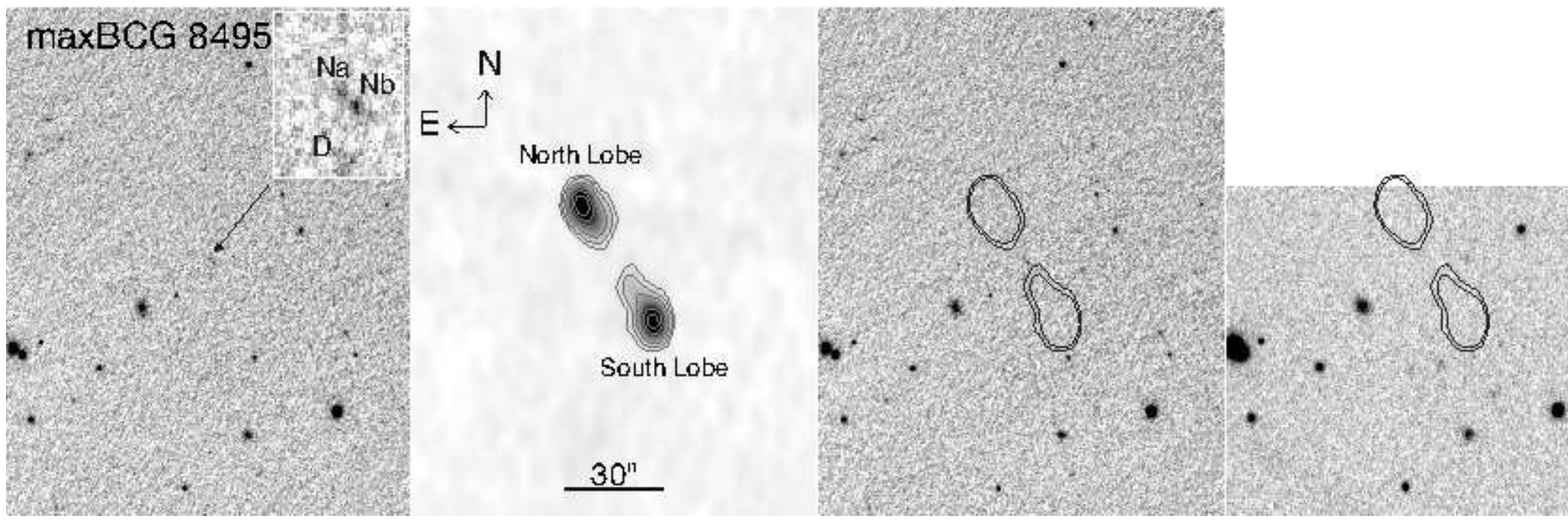}

\caption{From left to right, inverted greyscale image of the NIR
emission, FIRST greyscale emission with superimposed radio contours,
the first two levels of the FIRST contour plots superimposed on the NIR field and
likewise superimposed on the SDSS
field and the Stripe 82 (maxBCG 6167 and 8495) $r_{AB}$-band field. The field numbers are the sequential numbers of SDSS macxBCG cluster
fields to which the radio sources belong as they appear in Koester et al. (2007), letters/numbers are
the NIR components and lobe designations refer to the radio components. The inlays contain the NIR N component (Table~3 and 4), coincident
with the center of the radio structure. Radio contours are: (top) 0.0004, 0.0008, 0.0016 mJy (in log);
(bottom) 0.0005, 0.0010, 0.0020, 0.0040, 0.0060, 0.0080 mJy.}
\label{6167_8495}

\end{figure*}
\normalsize


\clearpage


\begin{figure*}
\centering
\includegraphics[width=8cm]{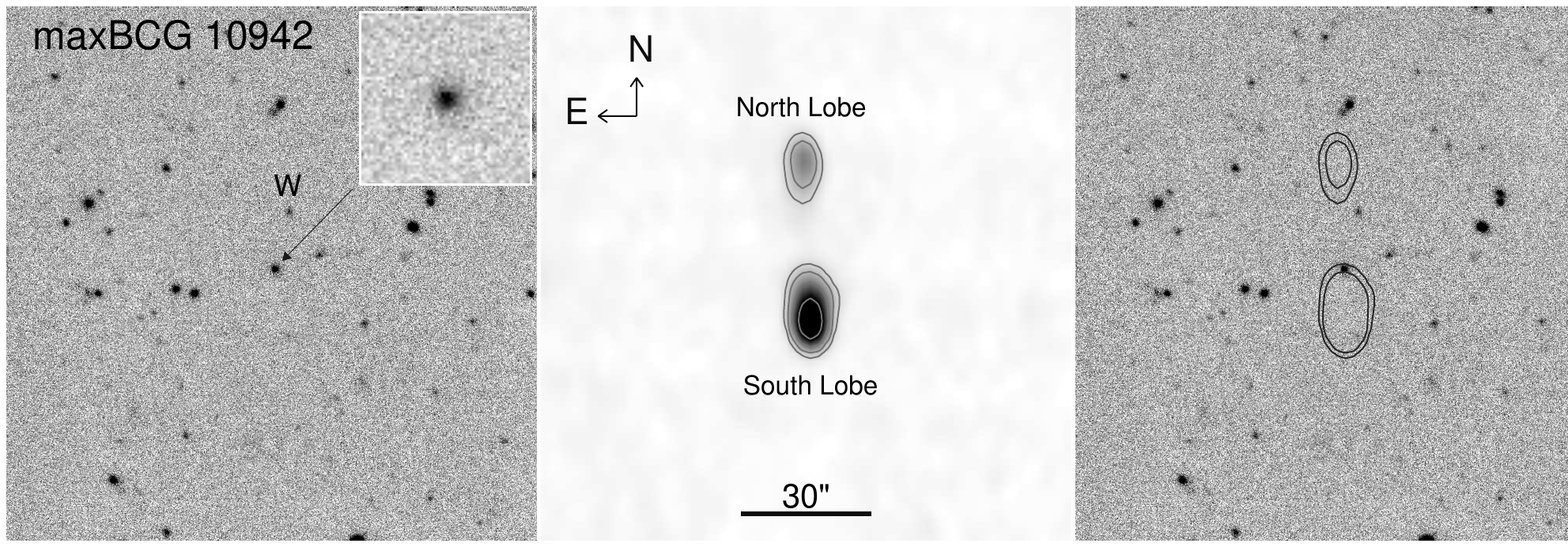}

\includegraphics[width=8cm]{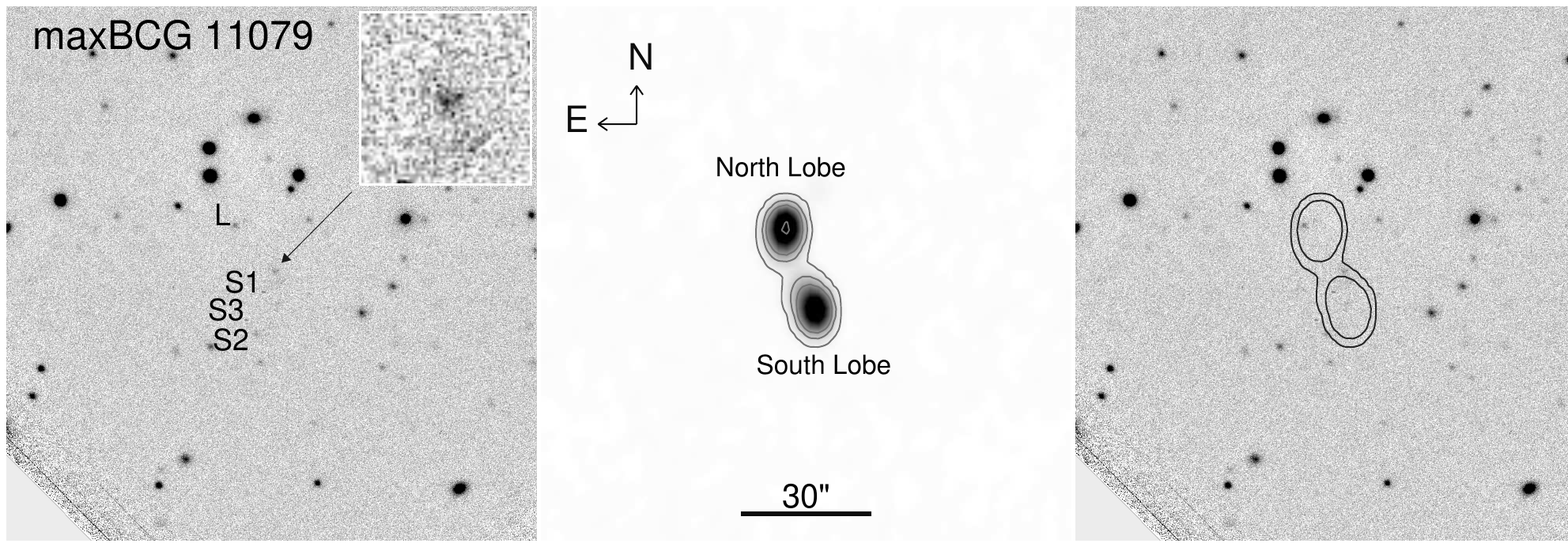}

\caption{From left to right, inverted greyscale image of the NIR
emission, FIRST greyscale emission with superimposed radio contours,
the first two levels of the FIRST contour plots superimposed on the NIR field and
likewise superimposed on the SDSS $r_{AB}$-band
field. The field numbers are the sequential numbers of SDSS macxBCG cluster
fields to which the radio sources belong, as they appear in Koester et al. (2007), letters/numbers are
the NIR components and lobe designations refer to the radio components. The inlays contain the NIR N component (Table~3 and 4), coincident
with the center of the radio structure. Radio contours are: (top) 0.001, 0.002, 0.004, 0.008,
0.016 mJy; (bottom) 0.008, 0.0016, 0.0032, 0.0064, 0.0128, 0.0256, 0.0512
mJy.}
\label{10942_11079}
\end{figure*}
\normalsize


\clearpage


\begin{figure*}
\centering
\includegraphics[width=8cm]{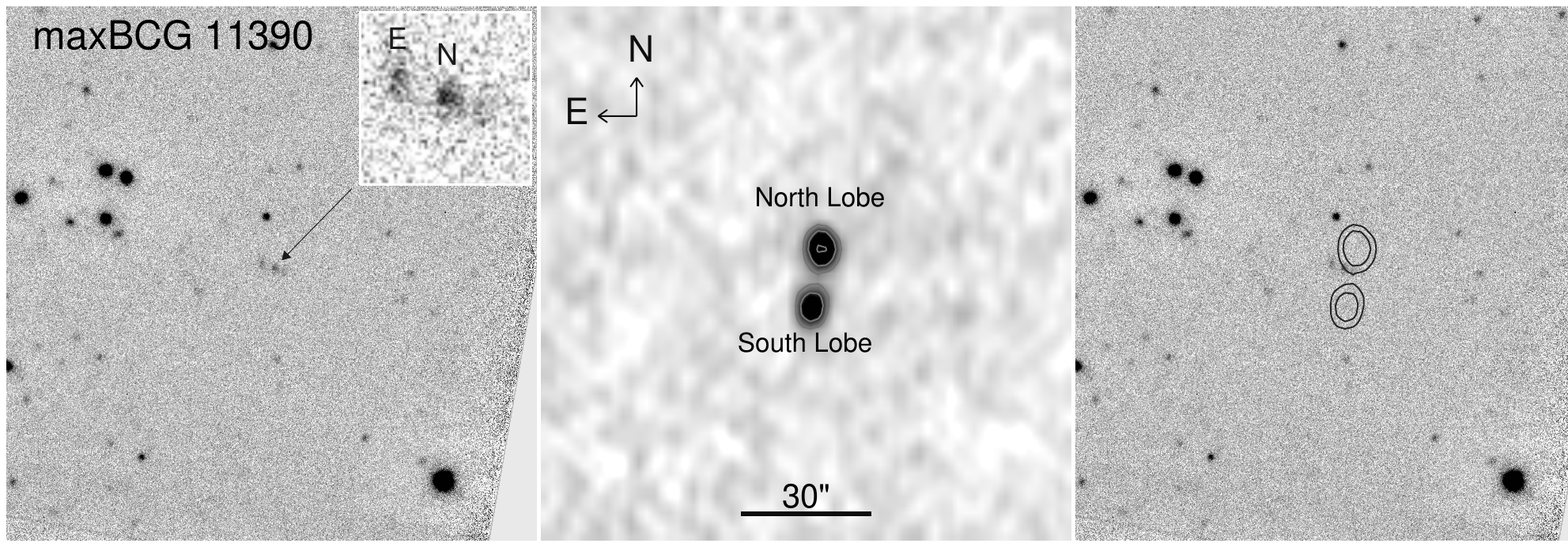}

\includegraphics[width=8cm]{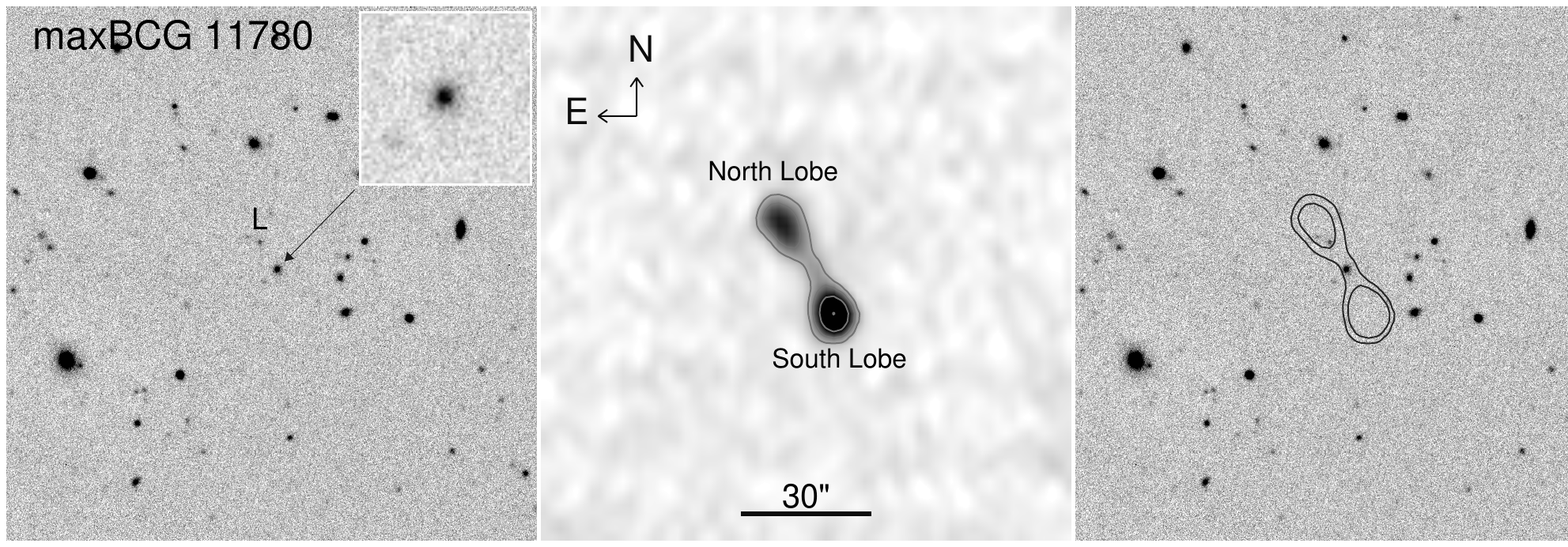}

\caption{From left to right, inverted greyscale image of the NIR emission,
FIRST greyscale emission with superimposed radio contours,
the first two levels of the FIRST contour plots superimposed on the NIR field and
likewise superimposed on the SDSS $r_{AB}$-band
field.  The field numbers are the sequential numbers of SDSS macxBCG cluster
fields to which the radio sources belong, as they appear in Koester et al. (2007), letters/numbers are
the NIR components and lobe designations refer to the radio components. The inlays contain the NIR N component (Table~3 and 4), coincident
with the center of the radio structure. Radio contours are: (top) 0.0004, 0.0008, 0.0016,
0.0032 mJy (in log); (bottom) 0.0005, 0.0010, 0.0020, 0.0040, 0.0080 mJy.}
\label{11390_11780}
\end{figure*}
\normalsize


\clearpage

\setcounter{figure}{4}


\begin{figure}
\centering
\includegraphics[width=6cm]{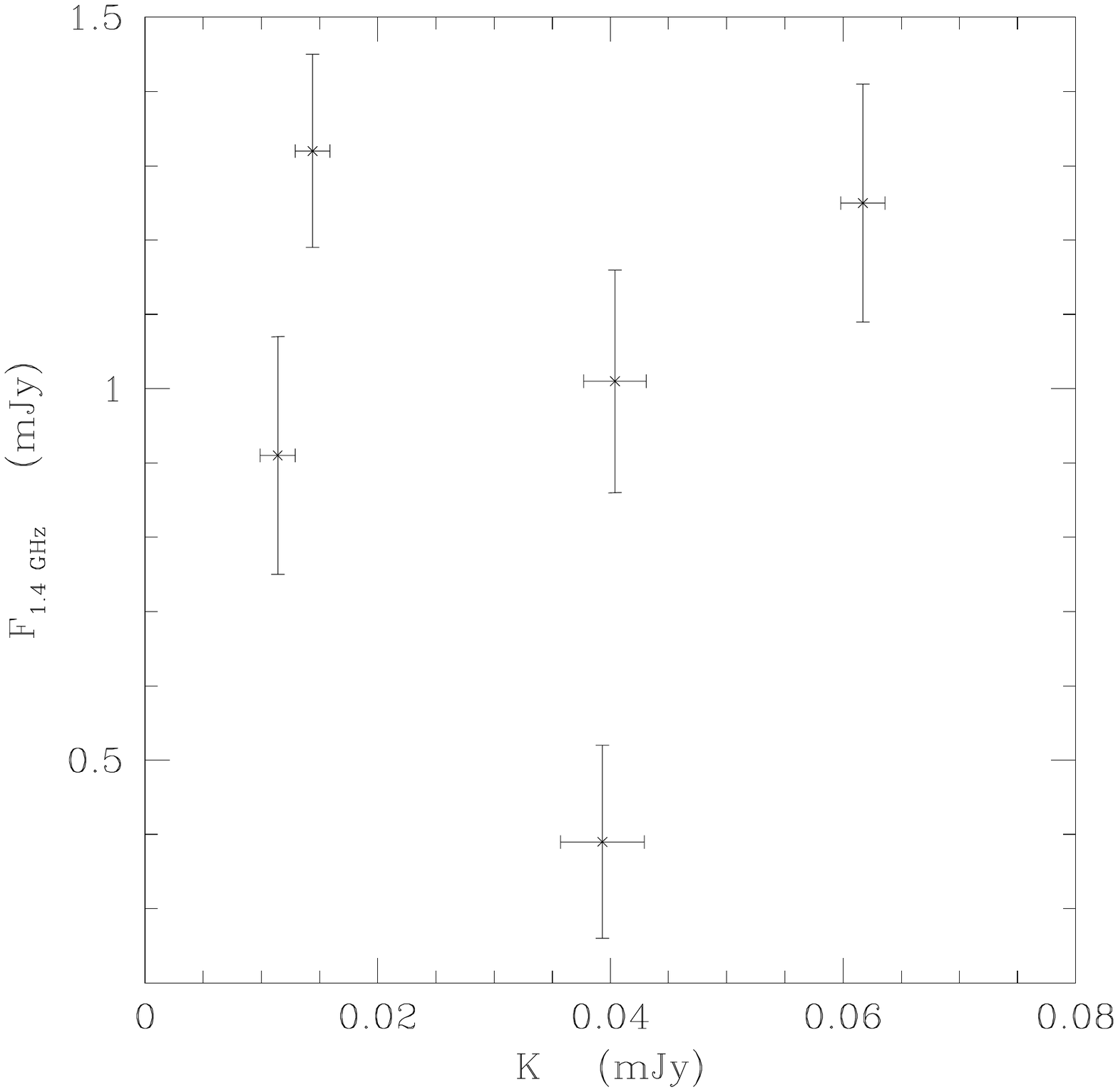}
\includegraphics[width=6cm]{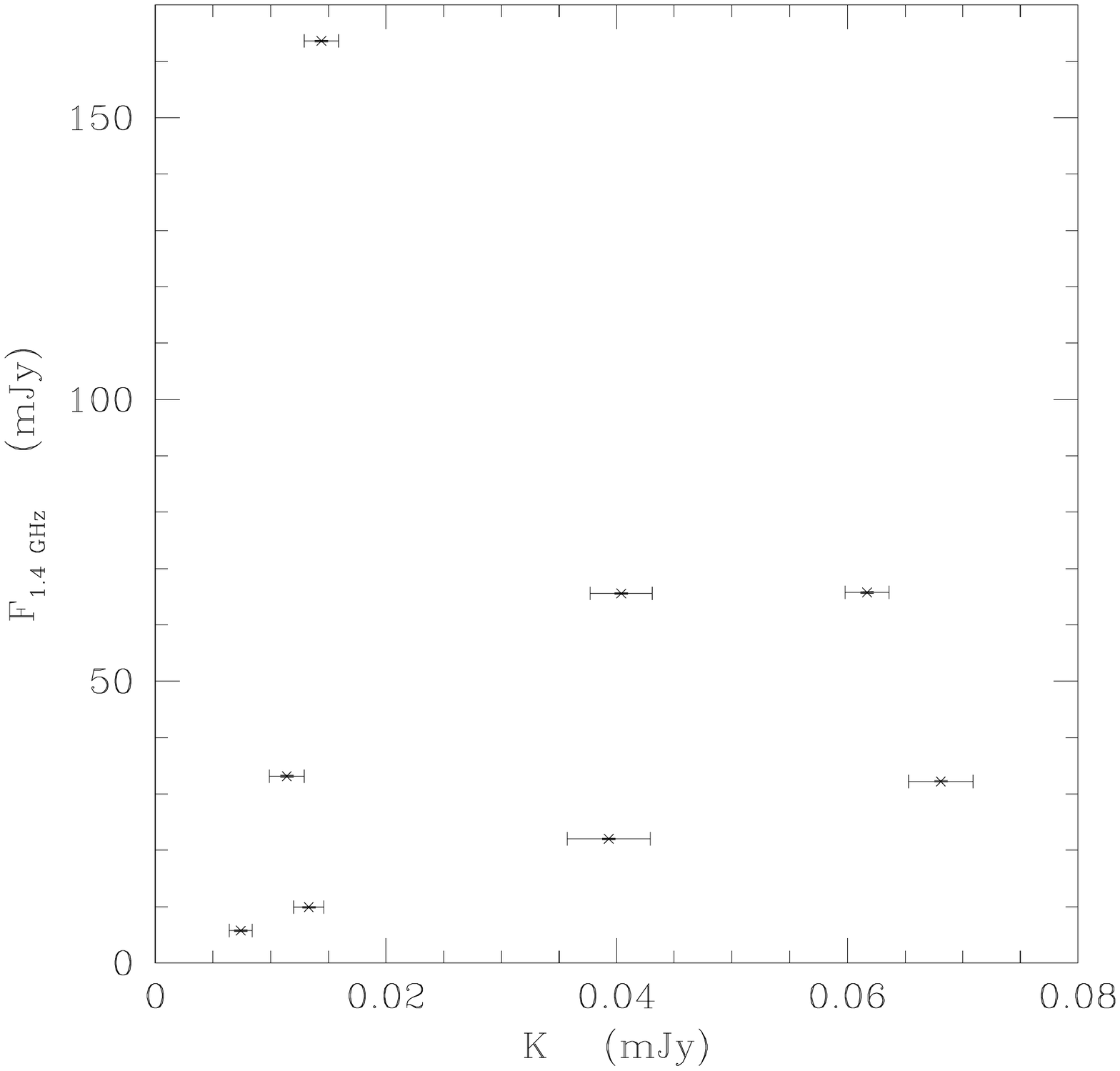}
\caption{The NIR versus 1.4 GHz FIRST flux density. The NIR flux pertains to the $K_s$-band fluxes of the NIR components (components N; Table~3) 
coincident with the radio structure centers. The NIR magnitudes are in the Vega system. The radio core flux density (top) pertains to the radio "core" flux density (Table~6) estimated using {\sc funtools} 
on circular regions of approximately the same size ($\sim$2~arcsec) and centered
on the NIR emission regions N (Table~3; see Section 5.2). The total radio flux density (bottom) pertains to the total FIRST flux density as a sum of the flux density of all the FIRST components 
(Table~1). The
errorbar associated with the FIRST flux density corresponds to 1$\times rms$, where $rms$ is the root mean square, a measure of the noise level in the FIRST image.}

\end{figure}
\normalsize


\clearpage


\begin{figure}
\centering
\includegraphics[width=6cm]{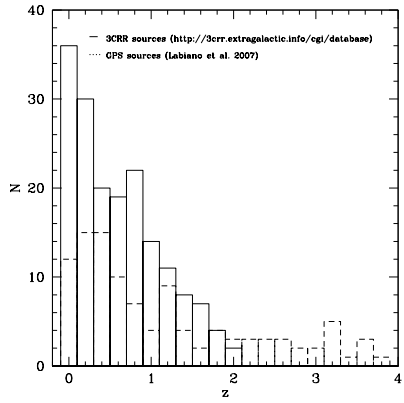}
\caption{The redshift distribution of the 3CRR (solid line) and GPS (dashed line) sample.}
\end{figure}
\normalsize



\begin{figure}
\centering
\includegraphics[width=6cm]{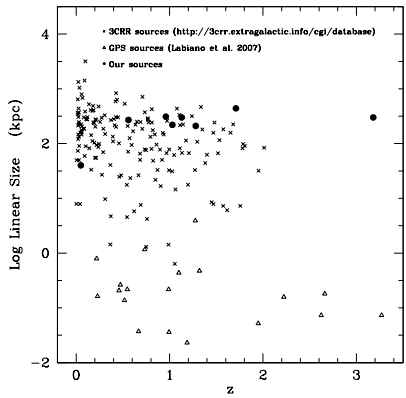}
\caption{The redshift versus largest linear size (LLS) 
for the 3CRR (crosses), GPS (triangles) and our sample (circles). The LLS for the GPS sources
were obtained from O'Dea \& Baum (1997), when a measurement was available.}
\end{figure}
\normalsize

\clearpage



\begin{figure*}
\includegraphics[width=6.0cm]{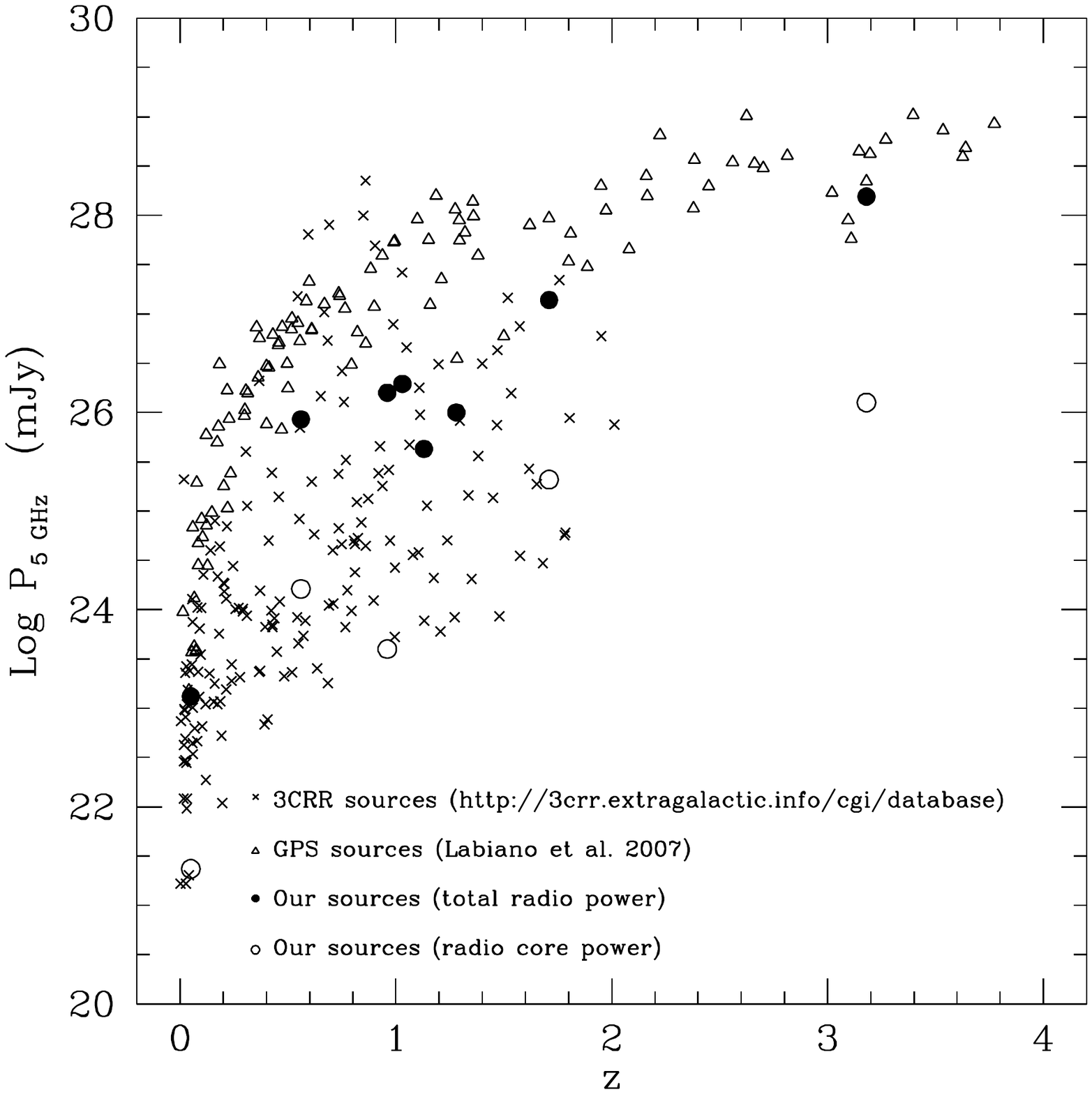}
\includegraphics[width=6.0cm]{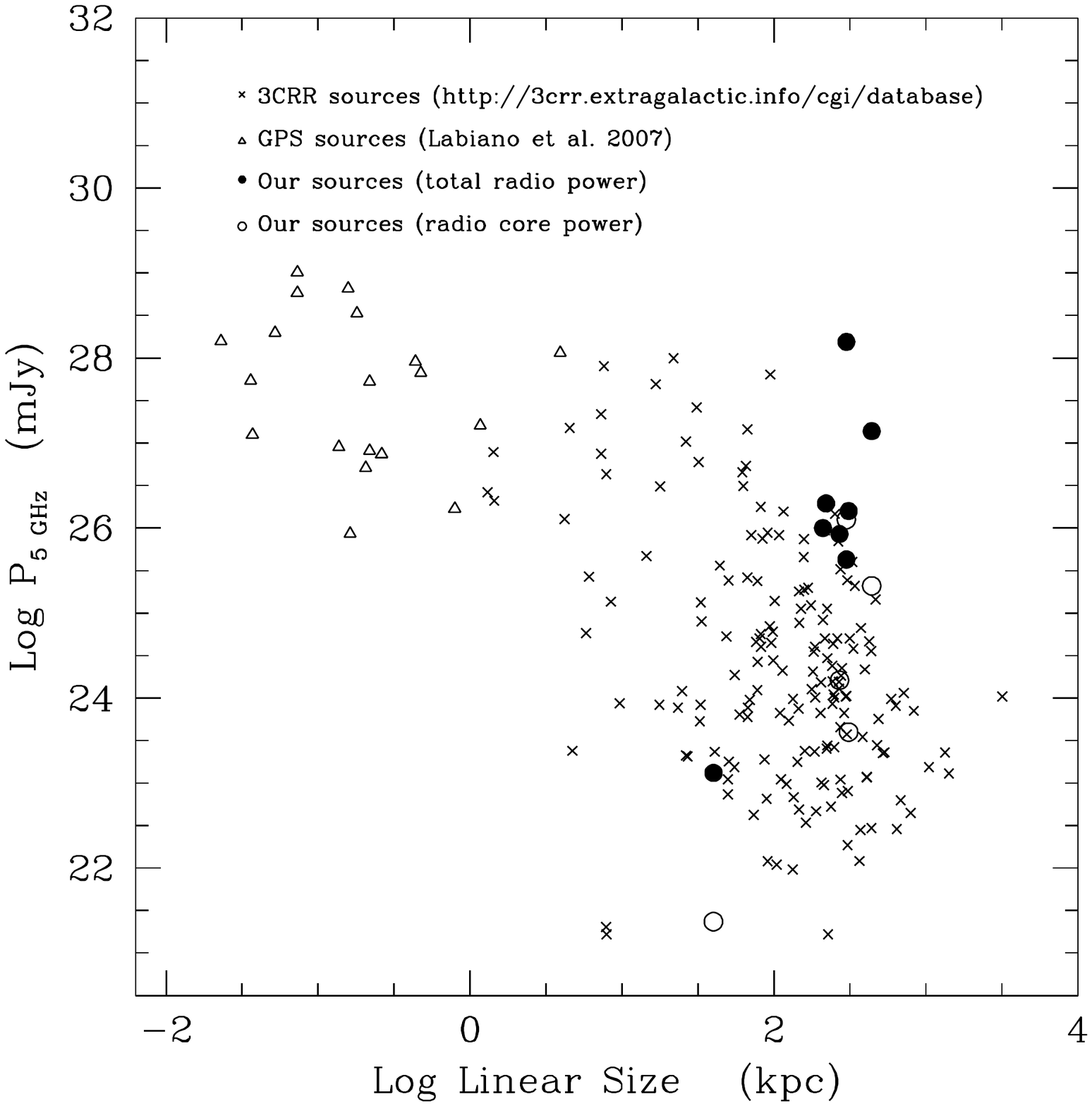}

\includegraphics[width=6.0cm]{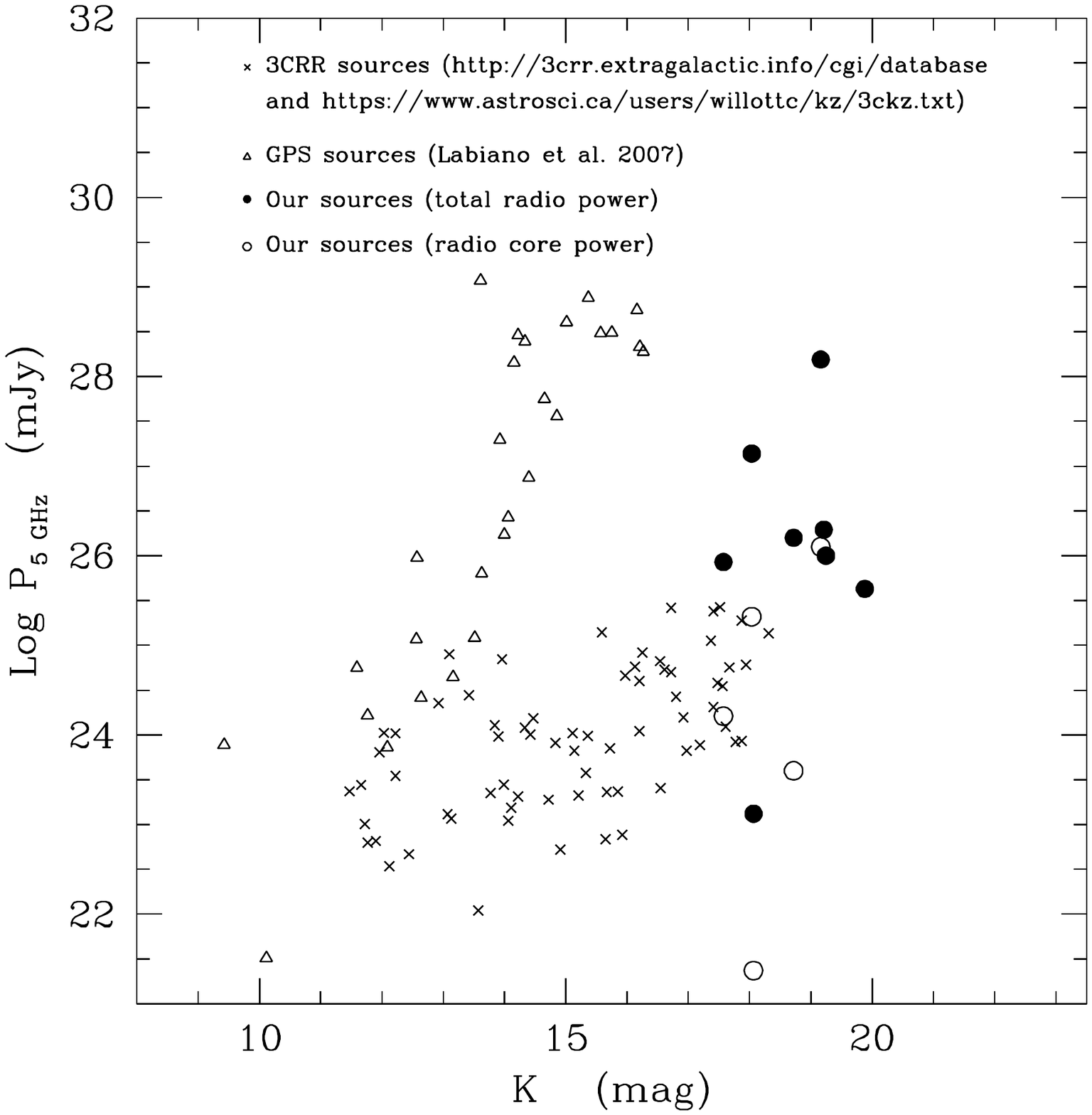}
\includegraphics[width=6.0cm]{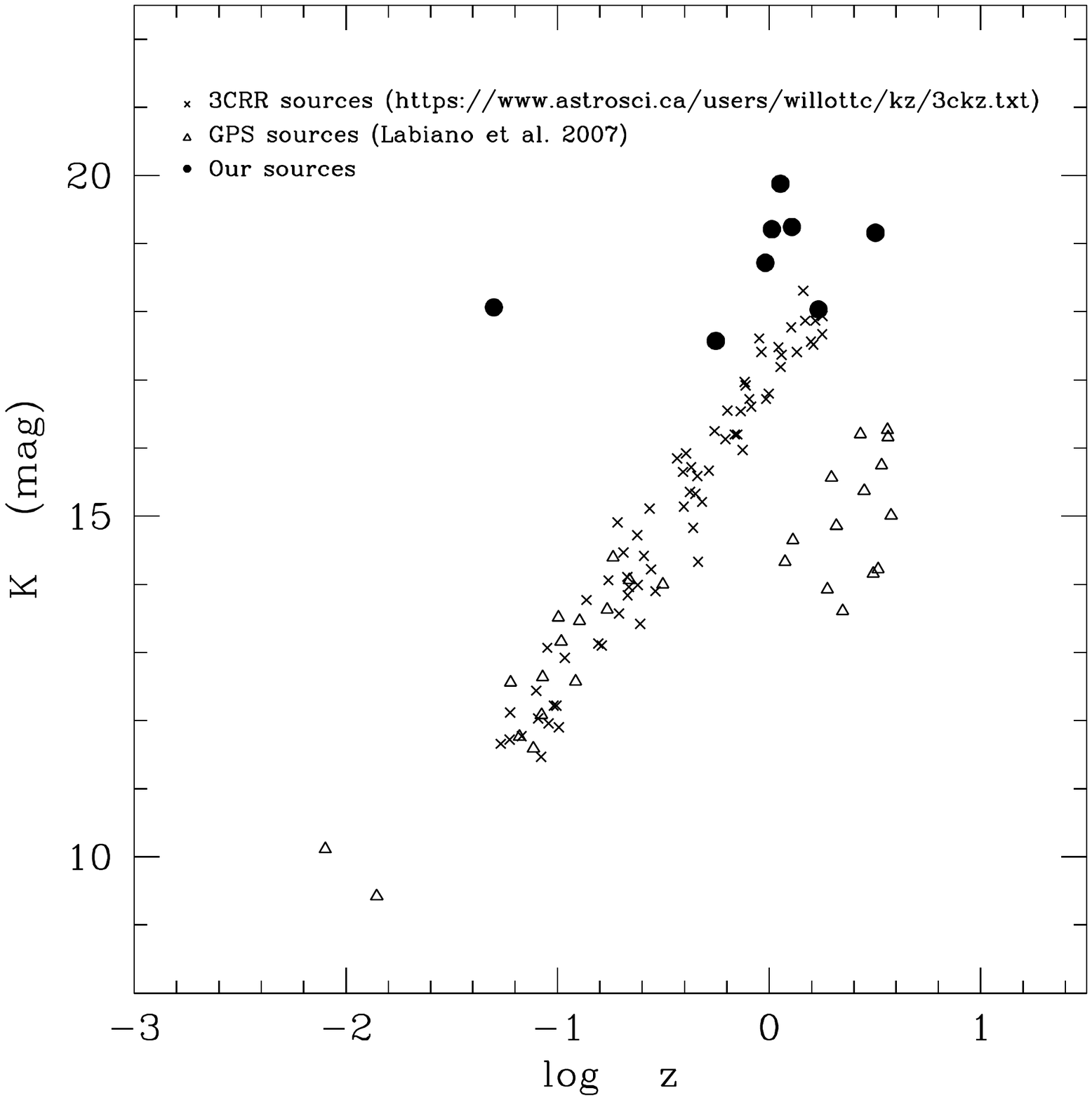}
\caption{Clockwise from the top left: The redshift versus the 5 GHz power for the 3CRR (crosses), GPS (triangles) and our sample total (closed circles) and core (open circles); the largest linear size (LLS) versus the 5 GHz power for the 3CRR (crosses), GPS (triangles) and our sample total (closed circles) and
core (open circles); the redshift versus
the $K$- (GPS and 3CRR) or $K_s$-band (our sample) magnitude for the 3CRR (crosses), GPS (triangles) and our sample (circles); 
the $K$- (GPS and 3CRR) or $K_s$-band (our sample) magnitude versus the 5 GHz power for the 3CRR (crosses), GPS (triangles) and our sample
total (closed circles) and core (open circles).
The radio measurements for the 3CRR sample are for the radio cores and 
for the GPS sample it is the total radio power. For our sample we include the radio "core" power (open circles; Table~7) 
estimated using {\sc funtools} on circular regions of approximately the same size ($\sim$2~arcsec) and centered
on the NIR emission regions N (Table~3; see Section 5.2) and the total radio power (closed circles; Table~7), the sum of the power of all the FIRST components 
(Table~1). We have assumed a flat radio spectrum for all sources in order to normalize to 5 GHz. The LLS for the GPS sample where obtained from O'Dea \& Baum (1997), when a measurement was available.
The 3CRR $K$-band magnitudes are aperture- and emission-line-corrected measurements for a subset of 3CRR sources.
For the GPS sample, the $K$-band magnitudes were obtained by interpolation from NED NIR photometric points. NIR magnitudes for our sample
are for the central NIR component, coincident with the center of the radio structure (components N; Table~3). 
The NIR magnitude system is Vega.}
\end{figure*}
\normalsize


\clearpage

\begin{figure*}
\centering
\includegraphics[width=10cm]{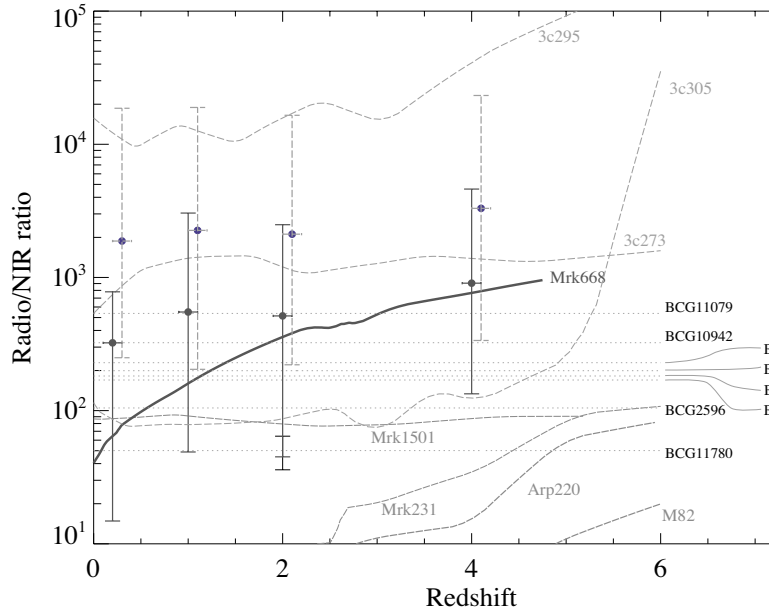}
\caption{The radio-NIR ratio as a function of redshift for different galaxy templates (solid and dashed lines). The dotted
lines are the radio-NIR ratios of our sample. The radio flux density of the optically faint radio sources 
pertains to the total FIRST radio flux density, from the sum of the integrated power of all the FIRST components (Table~1) and the NIR emission pertains to
the central NIR component, coincident with the center of the radio structure (components N; Table~3). The NIR magnitude system is Vega.
The large dots are the 
median redshift values for the 3CR (dashed; Spinrad 1985) and GPS (solid; Labiano \etal 2007) sample sources, with a one sigma errorbar. }
\end{figure*}
\normalsize


\section{Acknowledgments}

M.~E.~Filho acknowledges support from the Funda\c c\~ao para a Ci\^encia e 
Tecnologia (FCT), Minist\'erio da Ci\^encia e Ensino Superior, Portugal through the
grant SFRH/BPD/36141/2007. S. Ant\'on acknowledges FCT through the contract Ci\^encia2007.
All authors acknowledge support from the FCT
through the grant PTDC/CTE-AST/66147/2006 for the project "Understanding Massive Galaxies in the Universe".
We would like to thank the anonymous referee for his useful suggestions.

This research has made use of
NED (NASA/IPAC Extragalactic Database),
which is operated by the Jet Propulsion Laboratory, California Institute
of Technology, under contract with the National Aeronautics and Space
Administration. 

Funding for the SDSS and SDSS-II has been provided by the Alfred P. Sloan Foundation, the Participating Institutions, 
the National Science Foundation, the U.S. Department of Energy, the National Aeronautics and Space Administration, the Japanese Monbukagakusho, 
the Max Planck Society, and the Higher Education Funding Council for England. The SDSS Web Site is http://www.sdss.org/.

The SDSS is managed by the Astrophysical Research Consortium for the Participating Institutions. The Participating Institutions are the 
American Museum of Natural History, Astrophysical Institute Potsdam, University of Basel, University of Cambridge, Case Western Reserve University, 
University of Chicago, Drexel University, Fermilab, the Institute for Advanced Study, the Japan Participation Group, Johns Hopkins University, the Joint 
Institute for Nuclear Astrophysics, the Kavli Institute for Particle Astrophysics and Cosmology, the Korean Scientist Group, the Chinese Academy of Sciences 
(LAMOST), Los Alamos National Laboratory, the Max-Planck-Institute for Astronomy (MPIA), the Max-Planck-Institute for Astrophysics (MPA), New Mexico State 
University, Ohio State University, University of Pittsburgh, University of Portsmouth, Princeton University, the United States Naval Observatory and the 
University of Washington.


{}

\end{document}